\documentclass[12pt]{iopart}
\usepackage{iopams}

\usepackage{amssymb}
\usepackage{graphicx}
\usepackage{epsfig}
\usepackage{color}
\usepackage{url}
\usepackage{times}
\usepackage{bm}
\usepackage{mathrsfs}
\usepackage[utf8]{inputenc}
\usepackage{hyperref}
\usepackage{enumerate}
\usepackage{amsthm}
\usepackage{verbatim}
\usepackage{cite}
\usepackage{bbm}
\usepackage{stmaryrd}
\usepackage{slashed}
\usepackage{upgreek}
\usepackage{bbold}

\usepackage{caption}

\newcommand{\beq}{\begin{equation}}
\newcommand{\eeq}{\end{equation}}
\newcommand{\bea}{\begin{eqnarray}}
\newcommand{\eea}{\end{eqnarray}}
\newcommand{\bit}{\begin{itemize}}
\newcommand{\eit}{\end{itemize}}
\newcommand{\ben}{\begin{enumerate}}
\newcommand{\een}{\end{enumerate}}
\newcommand{\nn}{\nonumber}
\newcommand{\dfrac}[2]{{\displaystyle\frac{#1}{#2}}}
\newcommand{\eqref}[1]{(\ref{#1})}

\def\scri{\mathscr{I}}

\newcommand{\rh}{r_{\rm h}}
\newcommand{\rhi}{r_{{\rm h}_i}}
\newcommand{\rhj}{r_{{\rm h}_j}}
\newcommand{\sigmah}{\sigma_{\rm h}}
\newcommand{\sigmahi}{\sigma_{{\rm h}_i}}
\newcommand{\sigmai}{\sigma_{\rm i}}
\newcommand{\sigmaf}{\sigma_{\rm f}}
\newcommand{\io}{\rm in\mbox{-}out}
\newcommand{\oi}{\rm out\mbox{-}in}

  \newcommand{\NBI}{\address{$^{1}$Niels Bohr International Academy, Niels Bohr Institute, Blegdamsvej 17, 2100 Copenhagen, Denmark}}
          
\theoremstyle{plain}

\newcounter{mnotecount}

\newcommand{\mnotex}[1]
{\protect{\stepcounter{mnotecount}}$^{\mbox{\footnotesize $\bullet$\themnotecount}}$ 
\marginpar{
\raggedright\tiny\em
$\!\!\!\!\!\!\,\bullet$\themnotecount: #1} }

\begin{document}
\title{Hyperboloidal approach for static spherically symmetric spacetimes: a didactical introduction and applications in black-hole physics}
\author{Rodrigo Panosso Macedo}
\NBI
\eads{\mailto{rodrigo.macedo@nbi.ku.dk}}
\date{\today}

\begin{abstract}
This work offers a didactical introduction to the calculations and geometrical properties of a static, spherically symmetric spacetime foliated by hyperboloidal time surfaces. We discuss the various degrees of freedom involved, namely the height function, responsible for introducing the hyperboloidal time coordinate, and a radial compactification function. A central outcome is the expression of the Trautman-Bondi mass in terms of the hyperboloidal metric functions. Moreover, we apply this formalism to a class of wave equations commonly used in black-hole perturbation theory. Additionally, we provide a comprehensive derivation of the hyperboloidal minimal gauge, introducing two alternative approaches within this conceptual framework: the in-out and out-in strategies. Specifically, we demonstrate that the height function in the in-out strategy follows from the well-known tortoise coordinate by changing the sign of the terms that become singular at future null infinity. Similarly, for the out-in strategy, a sign change also occurs in the tortoise coordinate's regular terms. We apply the methodology to the following spacetimes: Singularity-approaching slices in Schwarzschild, higher-dimensional black holes, black hole with matter halo, and Reissner-Nordstr\"om-de Sitter. From this heuristic study, we conjecture that the out-in strategy is best adapted for black hole geometries that account for environmental or effective quantum effects.
 \end{abstract}

\maketitle
\pagestyle{plain} 
\section{Introduction}
In 1963, Penrose's seminal work on the ``Conformal Treatment of Infinity"\cite{Penrose:1964ge,Penrose:1964Republication} introduced a powerful tool to studies on the asymptotic behaviour of gravitational fields. The use of conformal methods in general relativity also enhanced our comprehension of black hole horizons and their geometrical properties. Apart from offering new perspectives on the nature of spacetime singularities, the formalism also introduced the concept of conformal (Carter-Penrose) diagrams, which provides an intuitive visualisation for the causal structure of black hole spacetimes.

This work considers the contribution of the conformal approach in general relativity to theoretical studies underpinning gravitational wave astronomy. We focus on scenarios modelled by black-hole perturbation theory. In particular, we are interested in two stages during the dynamical evolution of binary black holes: (i) the inspiral of extreme mass-ratios (EMRI) binaries and (ii) the after merger evolution of a deformed black hole. The former constitutes one of the main sources of gravitational wave for the future space antenna LISA~\cite{LISA}, with the dynamics of EMRIs rigorously modelled by the gravitational self-force programme \cite{Barack:2018yvs,Pound:2021qin}. The latter underlies the black-hole spectroscopy programme \cite{Dreyer:2003bv,Berti:2005ys,Giesler:2019uxc,Cabero:2019zyt,Baibhav:2023clw}, for which the notion of quasi-normal modes \cite{Chandrasekhar:579245,Kokkotas99a,Nollert99,Berti:2009kk,Konoplya:2011qq} plays a crucial role.

For astrophysics and gravitational wave physics, two particular null hyper-surfaces are of fundamental importance: the black-hole horizon ${\cal H}^+$ and future null infinity $\scri^+$ (or a cosmological horizon $r_{\Lambda}$). Despite the elegant abstract geometrical description of spacetimes given by Einstein's theory and the conformal approach to general relativity, the theoretical and numerical calculations underlying the study of astrophysical black holes and gravitational wave astronomy heavily rely on the use of coordinate systems. In the context, a natural foliation of the conformal spacetime arises in terms of the so-called hyperboloidal surfaces\cite{Friedrich1983,Frauendiener:2000mk,Zenginoglu:2007it}. Hyperboloidal foliations are spacelike hypersurfaces with asymptotically hyperbolic geometry. Intuitively, one possible visualisation for a hyperboloidal slice are horizontal lines in a Carter-Penrose diagram that crosses ${\cal H}^+$ and $\scri^+$.

In perturbation theory, the idea of adjusting the coordinate system to exploit the causal structure of asymptotic regions goes back to ref.~\cite{Schmi93}. However, it was only in the past decade that the hyperboloidal approach became a central framework to black hole perturbation theory. Indeed, ref.~\cite{Zenginoglu:2011jz} points out that the framework resolves problems regarding the representation of quasinormal mode eigenfunctions, whereas ref.~\cite{jaramillo2021pseudospectrum} imports the notion of pseudospectra into gravity by identifying that the hyperboloidal approach casts black-hole perturbation theory in terms of the spectral problem of a non-self adjoint operator\cite{Trefethen:2005,Sjostrand2019,dyatlov2019mathematical,Ashida:2020dkc}.

One of the main contribution from the hyperboloidal framework is the treatment of boundary conditions to the wave equations underlying the physical problems. Traditionally, black-hole perturbation theory is formulated in terms of coordinates $(t,r_*,\theta,\varphi)$ which closely resemble familiar coordinates in flat spacetime\footnote{The coordinate $r_* \in(-\infty, \infty)$ is usually referred as the tortoise coordinate in standard textbooks (see e.g.\cite{Misner:1973prb, Carroll:2004st} ),  it is closely related to spherical-like radial coordinate $r$ as defined in the upcoming sec.~\ref{sec:SchwarzschildCoord}}. Well-Posedness follows after boundary conditions are specified at the asymptotic region $r_*\rightarrow \infty$ and at the horizon $r_*\rightarrow -\infty$. However, along $t=$constant the limits $r_*\rightarrow \infty$ and $r_*\rightarrow -\infty$ correspond to spatial infinity $i^0$, and the bifurcation sphere ${\cal B}$. Therefore, one must impose external boundary conditions ensuring that the energy is absorbed by the black hole and propagates out to the wave zone to model the relevant physical scenario. Accessing these infinitely far regions is not straightforward from a numerical perspective. In practical terms, the numerical grid is cut at finite radius $r_*=\pm \,  r_*^{\rm cut}$, where the relevant boundary conditions are approximated. Then, the systematic errors are controlled by the location of this artificial cut, and the information about the solution for $r_*>\pm \, r_*^{\rm cut}$ is inferred or extrapolated from the values at finite radius. Besides, functions associated to quasi-normal modes diverge at $i^0$ and ${\cal B}$, which prevents the rigorous interpretation of the quasi-normal modes as a formal eigenvalue problem in an appropriated Hilbert space\cite{Kokkotas99a,Nollert99}.

This strategy, however, has reached its limits. High precision gravitational astronomy requires information from perturbation theory beyond the leading order\cite{Brizuela:2009qd,Loutrel:2020wbw,Miller:2020bft}. The dynamics of second order perturbation fields are sourced by the solution at first order, and thus they are dictated by the global behaviour of first order solutions. Any type of systematic error arising from the numerical solutions at first order will accumulate quadratically and jeopardise the accuracy for the second order studies. Besides, rigorous studies on the field equations shows that the second order sources have a better regularity behaviour when the system is parametrised by coordinates adapted to causal structure of the wave zone and the black hole\cite{Miller:2020bft}.

The hyperboloidal approach resolves the boundary condition issues when source terms are regular in the asymptotic regions. The application of the framework in black-hole perturbation theory has reached a mature stage, and it has established itself as an essential method to the study of wave propagation on a fixed background. Initially, the works focused on the development of numerical codes for time evolutions, benchmarked by the study of the late time decay of several fields propagating in black-hole spacetimes~\cite{Zenginoglu:2008wc,Zenginoglu:2008uc,Zenginoglu:2009hd,Zenginoglu:2009ey,Bizon:2010mp,Cruz-Osorio:2010rsr,Zenginoglu:2010zm,Zenginoglu:2010cq,Racz:2011qu,Jasiulek:2011ce,Harms:2013ib,Yang:2013uba,Spilhaus:2013zqa,PanossoMacedo:2014dnr,Hilditch:2016xzh,Csukas:2019kcb,OBoyle:2022yhp,Csukas:2021sia}.  In this context, the hyperboloidal foliations also offers the correct tool for rigours mathematical statements about the perturbations' late time decay and black-hole dynamical stability \cite{Andersson:2019dwi,Angelopoulos:2021hlw,Angelopoulos:2021cpg,Gajic:2022czq,Gajic:2022pst}.

Complementary to works in the time domain, refs.~\cite{Ansorg:2016ztf,PanossoMacedo:2018hab,PanossoMacedo:2019npm} established the use of hyperboloidal approach in the frequency domain (see also \cite{Ripley:2022ypi}). Of particular relevance was the identification of the so-called minimal gauge, which provides a geometrical understanding for the methods introduced by the seminal works of Leaver~\cite{Leaver85,Leaver90}. The hyperboloidal framework also allows to formally define quasi-normal modes as eigenvalues of the generator of time translations for a null foliation, acting on an appropriate Hilbert space\cite{Gajic:2019oem,Gajic:2019qdd}. Furthermore, the notion of quasinormal mode pseudospectra \cite{jaramillo2021pseudospectrum,jaramillo2021gravitational,Destounis:2021lum,Boyanov:2022ark,Sarkar:2023rhp,Arean:2023ejh} sheds light on the phenomenon of quasi-normal modes spectral instability~\cite{Aguirregabiria:1996zy,Vishveshwara:1996jgz,Nollert:1996rf}, which might impact predictions from the black-hole spectroscopy programme\cite{jaramillo2021gravitational,Cheung:2021bol,Berti:2022xfj,Kyutoku:2022gbr,Courty:2023rxk}. In the context of wave form modelling for EMRI's, the hyperboloidal approach enhanced the predictions arising from the effective-one-body approach~\cite{Bernuzzi:2010xj,Zenginoglu:2011zz,Bernuzzi:2011aj,Bernuzzi:2012ku,Harms:2014dqa,Nagar:2014kha,Harms:2015ixa,Harms:2016ctx,Lukes-Gerakopoulos:2017vkj}, and the framework has also excelled first tests in the gravitational self-force programme~\cite{Zenginoglu:2012xe,Wardell:2014kea,Thornburg:2016msc,PanossoMacedo:2022fdi,DaSilva:2023xif,Vishal:2023fye}.

By dwelling into the hyperboloidal framework for static, spherically symmetric spacetimes, the goal of this paper is to provide a didactical introduction to the topic so that students and interested researchers may familiarise themselves with the techniques currently employed in black-hole perturbation theory. In particular, eq.~\eqref{eq:BondiMass} presents an expression for the Trautman-Bondi mass in terms of generic hyperboloidal metric functions. Besides, eqs.~\eqref{eq:H_inout_flat} and \eqref{eq:Houtin} shows that hyperboloidal coordinates in the minimal gauge follows from a very simple algorithmic procedure: first, one constructs the well-known tortoise coordinate for the underlying spacetime; then the height function follows by a simple change in the sign of the terms with a singular behaviour at future null infinity.

The paper is organised as following. We introduce in the next section the hyperboloidal framework of a generic static, spherically symmetric spacetime. In particular, we provide a comprehensive discussion of the role played by all degrees of freedom and their geometrical significance. Sec.~\ref{sec:sphericalsymmetric} ends with a concrete example of a hyperboloidal foliation for the Schwarzschild solution. Then, sec.~\ref{sec:pert_theory} applies the framework to a class of wave equations commonly used in black-hole perturbation theory. Sec.~\ref{sec:min_gauge} offers a didactical exposition to the hyperboloidal minimal gauge. Finally, sec.~\ref{sec:examples_spacetime} applies the framework to several spacetimes. We employ geometrical units with the speed of light and Newton's gravitational constant $c=G=1$. Moreover, physical and conformal spacetimes are denoted by $({\cal M}, g_{ab})$ and $({\bar {\cal M}},\bar g_{ab} )$, respectively.

\section{Spherical Symmetric Spacetime}\label{sec:sphericalsymmetric}
\subsection{Schwarzschild coordinates $(t,r, \theta,\varphi)$}\label{sec:SchwarzschildCoord}
We first introduce the generic parametrisation for a spherically symmetric  line element in Schwarzschild-like coordinates $(t,r,\theta,\varphi)$ as
\beq
\label{eq:metric_tr}
ds^2 = - a(r) dt^2 + \dfrac{dr^2}{b(r)} + r^2 d\omega^2,
\eeq
with  $d\omega^2=\left( d\theta^2 + \sin^2\theta d\varphi^2\right)$ the line element of the $2$-sphere. This spherically symmetric line element decomposes the $4$-dimensional manifold ${\cal M}$ with coordinates $(t, r, \theta, \varphi)$ into two sub manifolds ${\cal M}= \overline{{\cal M}}^2 \times {\cal S}^2$. The $2$-dimensional Lorentzian manifold $ {\cal M}^2$ has coordinates $x^a=\{t, r\}$, whereas ${\cal S}^2$ represents a $2$-dimensional sphere with coordinates $x^{A}=\{\theta, \varphi\}$. 

The spacetime causal structure is captured by the behaviour of radial null geodesics. The radial character allows us to restrict ourselves to the Lorentzian manifold $ {\cal M}^2$, with the null geodesics parametrised by
\beq
\dfrac{dt}{dr} = \pm \dfrac{1}{f(r)}, \quad f(r) = \sqrt{a(r) b(r)}.
\eeq
Integrating the above equation yields
\beq
t = \pm r_*(r) + C
\eeq
with $C$ an arbitrary integration constant, and $r_*(r)$ the tortoise coordinate defined by
\beq
\label{eq:def_tortoise}
\dfrac{dr_*}{dr} = \dfrac{1}{f(r)}.
\eeq
The level set fixed by different values of the integration constant permit us to introduce the null (Eddington-Finkelstein) coordinates
\beq
u = t - r_*, \quad v =  t + r_*
\eeq
characterising outgoing ($u=$constant) and ingoing ($v=$constant) null rays.

With eq.~\eqref{eq:def_tortoise}, an alternatively parametrisation to the line element eq.~\eqref{eq:metric_tr} in terms of the tortoise coordinate $r_*$ follows via
\beq
\label{eq:metric_tr*}
ds^2 =  a(r)\bigg(-  dt^2 + dr_*^2\bigg) + r^2 d\omega^2, \quad r=r(r_*).
\eeq
In the definition of the tortoise coordinate, the asymptotic behaviour of $f(r)$ as $r\rightarrow \infty$, as well in the vicinity of spacetime horizons is of particular relevance. We assume the function $f(r)$ has $n_{\rm h}=N_{\rm h}+1$ real positive roots $\rhi$ ($i=0\cdots N_{\rm h}$), which allows for a decomposition
\beq
\label{eq:func_f_horizons}
f(r) =  {K}(r) \prod_{i=0}^{N_{\rm h}} \left(1- \dfrac{\rhi}{r} \right).
\eeq
The positive roots represent horizons in the spacetime. Around a value $\rhj$, we may also introduce the representation
\beq
\label{eq:func_f_horizons_j}
f(r) =  {K_j}(r) \left(1- \dfrac{\rhj}{r} \right). 
\eeq
For asymptotically flat $4$ dimensional spacetimes, we assume the asymptotic expansion
\beq
\label{eq:f_asymp_Mink}
f(r) = 1 - \dfrac{2M}{r} + {\cal O}\left( r^{-2} \right),
\eeq 
with $M$ the spacetime ADM mass. For asymptotic (Anti-)de Sitter spacetime, on the other hand, one encounters
\beq
\label{eq:f_asymp_dS}
f(r) = -\dfrac{\Lambda}{3}r^2 + {\cal O}\left( r^{-2} \right),
\eeq 
with the cosmological constant $\Lambda$ assuming positive values for de Sitter, and negative for Anti-de Sitter spacetimes.

\subsection{Hyperboloidal Coordinates $(\tau,\sigma,\theta,\varphi)$}\label{sec:Hyp_coord}
We introduce compact hyperboloidal coordinates $(\tau, \sigma, \theta, \varphi)$ via the scri-fixing technique~\cite{Zenginoglu:2007jw}, together with the compactification strategy from ref.~\cite{PanossoMacedo:2018hab}
\beq
\label{eq:HypCoord}
t = \lambda \bigg(\tau-H(\sigma)\bigg), \quad r= \lambda \dfrac{\rho(\sigma)}{\sigma}.
\eeq
Here, $\lambda$ is a given length scale of the spacetime, and the compact radial coordinate is defined in the domain $\sigma\in[\sigmai,\sigmaf]$. For instance, in an asymptotically flat black hole spacetime, the exterior region has $\sigmai=0$ locating $\scri^+$ and a final coordinate value $\sigmaf=1$ locating the event horizon. Here, we leave the notation generic to be adapted according to the specific scenario under consideration.

As we will show in sec.~\ref{sec:conformal}, $\rho(\sigma)$ represents the areal radius in the conformal space. Therefore we assume $\rho(\sigma)>0$ \cite{PanossoMacedo:2018hab}. Motivated by the differential form $dr = -\lambda(\rho - \sigma \rho')/\sigma^2 \, d\sigma$ that follows from the radial transformation eq.~\eqref{eq:HypCoord}, ref.~\cite{PanossoMacedo:2018hab} defines 
\beq
\label{eq:beta}
\beta(\sigma) := \rho(\sigma) - \sigma \rho'(\sigma).
\eeq
Besides, the coordinate change \eqref{eq:HypCoord} implies the transformation in the tortoise coordinate via
\beq
\label{eq:def_x}
x(\sigma) = \dfrac{r_*(r(\sigma))}{\lambda}.
\eeq
With the help of eq.~\eqref{eq:def_tortoise}, the dimensionless tortoise coordinate $x(\sigma)$ follows from integrating the differential equation
\beq
\label{eq:dx_dsigma}
x'(\sigma) = - \dfrac{\beta(\sigma)}{\sigma^2 {\mathcal F(\sigma)}}, \quad {\mathcal F(\sigma)}= f(r(\sigma)).
\eeq
As expected from eq.~\eqref{eq:func_f_horizons}, ${\mathcal F(\sigma)}$ vanishes at the horizons $\sigmahi$ defined by eq.~\eqref{eq:HypCoord} as $\rhi=r(\sigmahi)$. Indeed, the representations \eqref{eq:func_f_horizons} or \eqref{eq:func_f_horizons_j} yield respectvely
\bea
\label{eq:F_of_sigma}
{\mathcal F(\sigma)} &=& {\cal K}(\sigma) \prod_{i=0}^{N_{\rm h}} \left(1- \dfrac{\rhi}{r(\sigma)} \right), \quad {\cal K}(\sigma) = K(r(\sigma)),\\
\label{eq:F_of_sigma_j}
{\mathcal F(\sigma)} 	&=& {\cal K}_j(\sigma) \left(1- \dfrac{\rhj}{r(\sigma)} \right), \quad  {\cal K}_j(\sigma) = K_j(r(\sigma)).
\eea
For asymptotically flat spacetimes one reads from eq.~\eqref{eq:f_asymp_Mink} the behaviour
\beq
\label{eq:F_of_sigma_asymp_Mink}
{\mathcal F(\sigma)} = 1 - \dfrac{2M}{\lambda} \dfrac{\sigma}{\rho(0)} + {\cal O}(\sigma^2).
\eeq
For asymptotically (Anti-)de Sitter spacetimes one reads from eq.~\eqref{eq:f_asymp_dS} the behaviour
\beq
\label{eq:F_of_sigma_asymp_dS}
{\mathcal F(\sigma)} = -\dfrac{\lambda^2 \Lambda}{3}  \dfrac{\rho(0)^2}{\sigma^2}\bigg( 1 + {\cal O}\left(\sigma^{-1}\right)\bigg).
\eeq

The height function $H(\sigma)$ must ensure that the hypersurfaces $\tau=$ constant penetrate the black-hole horizon at $\sigmah$ and that it intersects future null infinity at $\sigma=0$. These conditions are discussed in ref.~\cite{Zenginoglu:2011jz}, and we will review them in the next sections according to the notation employed here. In particular, sec.~\ref{sec:min_gauge} will introduce choices for eqs.~\eqref{eq:HypCoord} within the so-called minimal gauge class.

The coordinate transformation \eqref{eq:HypCoord} yields the following spherical symmetric line element in hyperboloidal coordinates $(\tau,\sigma,\theta, \varphi)$ 
\beq
\label{eq:eq_hyp}
d s^2 =  \dfrac{\lambda^2}{\sigma^2}\, \Bigg\{  \zeta(\sigma) \beta(\sigma) \left[-p(\sigma) d\tau^2 + 2\gamma(\sigma)d\tau d\sigma + w(\sigma) d\sigma^2 \right] +\rho(\sigma)^2d\omega^2 \Bigg\}.
\eeq
Despite the rather cumbersome appearance, each one of the functions carries information either from the physical character of the spacetime, or the hyperboloidal gauge degrees of freedom.

\subsubsection{Conformal Transformation}\label{sec:conformal}
The first important property to observe is that the line element \eqref{eq:eq_hyp} assumes explicitly a conformal representation $ds^2 = \Omega^{-2} d\bar s^2$ in the hyperboloidal coordinates $(\tau, \sigma, \theta, \varphi)$ with
\beq
\label{eq:conf_metric}
d\bar s^2 =\Xi(\sigma)  \left[-p(\sigma) d\tau^2 + 2\gamma(\sigma)d\tau d\sigma + w(\sigma) d\sigma^2 \right] + \rho(\sigma)^2d\omega^2, \quad \Omega = \dfrac{\sigma}{\lambda}.
\eeq
This choice employs the conformal factor itself as the radial coordinate $\sigma = \Omega \lambda$. Thanks to this particular choice of conformal factor that takes into account the asymptotic properties of the metric, all functions $\Xi(\sigma)$, $p(\sigma)$, $\gamma(\sigma)$, $w(\sigma)$ and $\rho(\sigma)$ are regular in the domain $\sigma\in[\sigmai,\sigmaf]$. 

The next sections explain the interpretation of these functions in terms of the original physical degree of freedom in line element \eqref{eq:metric_tr} --- or alternatively eq.~\eqref{eq:metric_tr*} --- and the hyperboloidal gauge freedom when specifying the coordinate transformation \eqref{eq:HypCoord}. 

\subsubsection{The $(\tau,\sigma)$-sector of the conformal metric}
The transformation into the hyperboloidal coordinates \eqref{eq:HypCoord} respects the underlying spherical symmetry of the spacetime. Therefore, the $4$-dimensional conformal manifold $\overline{\cal M}$ also decomposes itself into two sub manifolds $\overline{\cal M}= \overline{{\cal M}}^2 \times {\cal S}^2$,  with coordinates $\bar x^a=\{\tau, \sigma\}$ for the Lorentzian manifold $\overline {\cal M}^2$, and $\bar x^{A}=\{\theta, \varphi\}$ for the sphere ${\cal S}^2$. The metric for $\overline{{\cal M}}^2$ reads
\bea
\left.d\bar s^2\right|_{\overline {\cal M}^2} &=& \bar\eta_{ab}dx^a dx^b \\
&=& \Xi(\sigma)  \left[-p(\sigma) d\tau^2 + 2\gamma(\sigma)d\tau d\sigma + w(\sigma) d\sigma^2 \right].
\eea
From the coordinate change leading to eq.~\eqref{eq:eq_hyp}, one identifies 
\beq
\label{eq:def_Xi_0}
\Xi(\sigma) = \zeta(\sigma) \beta(\sigma).
\eeq  
Anticipating a result from the next sections for the relation between $p(\sigma)$, $\gamma(\sigma)$ and $w(\sigma)$  --- c.f.~\eqref{eq:def_w} --- we can establish an alternative interpretation to $\Xi(\sigma)$ and define it as
\beq
\label{eq:def_Xi}
\Xi(\sigma) = \sqrt{ -\boldsymbol {\bar\eta}}, \quad \boldsymbol {\bar\eta} = \det{{\bar\eta_{ab}}}.
\eeq

\subsubsection{The $(\theta,\varphi)$-sector of the conformal metric}
The function $\rho(\sigma)$ introduced in eq.~\eqref{eq:HypCoord} relates to the gauge choice when compactifying the radial direction. As anticipated in sec.~\ref{sec:Hyp_coord}, $\rho(\sigma)$ is the areal radial function in the conformal metric. Indeed, the metric for $\bar{\cal S}^2$ reads
\bea
\left.d\bar s^2\right|_{\overline {\cal S}^2} &=& \bar q_{AB}dX^A dX^B \\
&=& \rho(\sigma)^2 \left( d\theta^2 + \sin^2\theta d\varphi^2\right).
\eea
Thus, we restrict ourselves to radial coordinate transformations \eqref{eq:HypCoord} with $\rho(\sigma)>0$.

\subsubsection{Interpretation of the metric components}


The physical properties of the spacetime are encoded in two degrees of freedom expressed either by $a(r)$ and $b(r)$ in eq.~\eqref{eq:metric_tr}, or $a(r)$ and $r_*(r)$ in eq.~\eqref{eq:metric_tr*}. In the hyperboloidal coordinates, these degrees of freedom are captured first by the function\footnote{Unless necessary, from now on we omit the explicit $\sigma$-dependence on the radial coordinate mapping $r(\sigma)$. Thus, an expression of the type $F(\sigma) = f(r)$ is understood as $F(\sigma) = f(r(\sigma))$ for generic functions $f$ and $F$. }
\beq
\label{eq:zeta}
\zeta(\sigma) = \sqrt{\dfrac{a(r)}{b(r)}},
\eeq
which measures the deviation between $g_{tt}$ and  $1/g_{rr}$ in eq.~\eqref{eq:metric_tr}. In particular, $\zeta(\sigma) \neq 1$ only for more intricate geometries. For asymptotically flat spacetimes, we can derive the asymptotic behaviour of $\zeta(\sigma)$ with the help from the Newtonian limit
\beq
\label{eq:zeta_asympot}
a(r)\sim 1 - \dfrac{\phi_{N}}{r},\quad b(r)^{-1}\sim 1 + \dfrac{\phi_{N}}{r} \Longrightarrow \zeta(\sigma) = 1 + {\cal O}(\sigma^2).
\eeq
In the hyperboloidal representation of the conformal metric \eqref{eq:conf_metric}, $\zeta(\sigma)$ follows from the definition \eqref{eq:def_Xi} via $\zeta(\sigma) = \Xi(\sigma)/\beta(\sigma)$.

The physical information carried originally by the tortoise coordinate $r_*(r)$ is now encoded in 
\beq
\label{eq:p_from_x}
p(\sigma) := -\lambda \dfrac{d\sigma}{dr_*}= -\dfrac{1}{x'(\sigma)}.
\eeq 
With the help of eq.~\eqref{eq:dx_dsigma}, it reads
 \beq
 \label{eq:def_p}
 p(\sigma)= \dfrac{\sigma^2 {\cal F}(\sigma)}{\beta(\sigma)}.
 \eeq
From eq.~\eqref{eq:F_of_sigma}, one notice that the roots of $p(\sigma)$ coincides with those of ${\cal F}(\sigma)$, i.e., the horizons of the spacetime. 
Besides, in asymptotically flat spacetimes, $p(\sigma)$ also vanishes at future null infinity $\sigma=0$. More specifically, considering eqs.~\eqref{eq:F_of_sigma} and \eqref{eq:F_of_sigma_asymp_Mink} it follows
\beq
\label{eq:prop_p}
p(\sigma) \sim \sigma^2 \prod_{i=0}^{N_{\rm h}}(\sigmahi - \sigma), 
\eeq 
i.e., $p(\sigma)$ vanishes quadratically at future null infinity ($\sigma=0$). Typically, $p(\sigma)$ vanishes linearly at the horizons, unless the spacetime contains an extremal black-hole, where the values of two horizons coincide.

The gauge freedom for the time transformation $t=t(\tau,\sigma)$ is encoded in the so-called boost-function~\cite{Zenginoglu:2011jz}
\bea
\label{eq:def_gamma}
\gamma(\sigma) := -\lambda \dfrac{dH}{dr_*} = H'(\sigma)p(\sigma).
\eea
The functions $\rho(\sigma), \zeta(\eta), p(\sigma)$ and $\gamma(\sigma)$ capture all degrees of freedom in the framework, i.e., two in the generic spherical symmetric line element \eqref{eq:metric_tr} and two from the the mapping $(t,r)\leftrightarrow(\tau,\sigma)$ in eq.~\eqref{eq:HypCoord}. Thus, the remaining function $w(\sigma)$ in the line element is not independent, but rather related to  $p(\sigma)$ and $\gamma(\sigma)$ via
\beq
\label{eq:def_w}
w(\sigma) := \dfrac{1-\gamma(\sigma)^2}{p(\sigma)}.
\eeq
Eq.~\eqref{eq:def_w} imposes constraints on the boost function $\gamma(\sigma)$, see ref.~\cite{Zenginoglu:2011jz}. The first arises from demanding $ g_{\sigma \sigma}>0$, which implies
\beq
\label{eq:spacelike}
\left\{
\begin{array}{ccc}
\left| \gamma \right| < 1 & {\rm if } & p(\sigma) > 0, \\
\\
\left| \gamma \right| > 1 & {\rm if } & p(\sigma) < 0. \\
\end{array}
\right.
\eeq
From the geometrical perspective, this condition ensures that the hypersurfaces $\tau=$ constant remain spacelike in the domains $\sigma\in[\sigmai,\sigmaf]$ as it will be demonstrated in sec.~\ref{sec:3+1}. The black-hole exterior region corresponds to $p(\sigma)>0$.

The second set of conditions follows from imposing regularity of $w(\sigma)$ at the roots of $p(\sigma)$, i.e., the horizons $\sigma=\sigmahi$, and $\sigma=0$ for asymptotically flat spacetimes. At the horizons, these conditions read
\beq
\label{eq:align_nullvectors_hrz}
\gamma(\sigma)^2 = 1 +{\cal O}(\sigmahi-\sigma) \quad \Longrightarrow \quad \gamma(\sigma) = \pm1 +{\cal O}(\sigmahi-\sigma).
\eeq
The choice for the positive or negative signal is discussed in the sec.~\ref{sec:nullvectors} as it implies having either the ingoing or an outgoing null vector as generator of the horizon surface.

For asymptotically flat spacetimes, $\gamma(\sigma)$ must behave as 
\bea
\label{eq:align_nullvectors_scri}
\gamma(\sigma)^2 = 1 +{\cal O}(\sigma^2) \quad &\Longrightarrow& \quad \gamma(\sigma) = 1 +{\cal O}(\sigma^2).
\eea
In this case, the particular choice for the plus sign, ensure that the vector $\partial_\tau$ is the generator of future null infinity. These geometrical interpretations will be elucidated in the next sections.

\subsubsection{3+1 representation}\label{sec:3+1}
The $3+1$ formalism in General Relativity allows one to rewrite the covariant Einstein's equations with focus on explicitly splitting the roles of space and time. The formalism is particular useful in numerical relativity as it allows the formulation of Einstein's equations as a Cauchy problem: given adequate initial and boundary conditions, the equations fully determine the future evolution of the system\cite{Gourgoulhon:2007ue,Alcubierre2007,Bona2009,Baumgarte2010,Baumgarte:2021skc}. The non-linear evolution of Einstein's equation in the hyperboloidal formalism has undergone great progress in the past years (see for instance refs.~\cite{Fodor:2003yg,Fodor:2006ue,Csizmadia:2009dm,Buchman:2009ew,Schinkel:2013zm,Maliborski:2017jyf,Gasperin:2019rjg,Frauendiener:2021eyv,Frauendiener:2022bkj,Duarte:2022vxn,Frauendiener:2023ltp,Vano-Vinuales:2023yzs} and references therein), but several challenges remain open (e.g. binary black-hole evolutions). Here, we identify from eq.~\eqref{eq:conf_metric} the basic quantities for a $3+1$ representation of the spacetime.

We begin by considering the form $\bar n_a = -\bar \nabla_a \tau$, which gives the (unnormalised) normal vector to the hypersurfaces $\tau =$ constant in the conformal spacetime \eqref{eq:conf_metric}
\beq
\label{eq:normal vector}
\bar n_a = -\delta_{a}^\tau, \quad \bar n^a = \dfrac{w}{\Xi} \delta^a_\tau -\dfrac{\gamma}{\Xi}\delta^a_\sigma, \quad || \bar n^a||^2 = - \dfrac{w}{\Xi}.
\eeq
The hypersurfaces $\tau =$ constant remain spacelike as long as $||\bar n^a||^2<0 \leftrightarrow w(\sigma)>0$, c.f.~eq.~\eqref{eq:spacelike}.

The three-dimensional metric $\gamma_{ij}$ ($i,j=1\cdots 3$) measuring proper distances in the slide $\tau=$constant has non-vanishing components
\beq
 \bar \upgamma_{\sigma \sigma} = \Xi w,  \quad  \bar \upgamma_{\varphi \varphi} = \dfrac{\upgamma_{\theta \theta}}{\sin^\theta} = \rho^2.
\eeq
The lapse function $\bar \alpha$ (measuring the lapse of proper time for an observer with $4-$velocity parallel to $\bar n^a$) and the shift vector $\bar \upbeta^i$ (measuring the relative velocity between this observer and the lines of constant spatial coordinate) are
\bea
\label{eq:3+1}
\bar \alpha^2 = \dfrac{\Xi}{w},  \quad \bar\upbeta_\sigma = \gamma \Xi,  \quad \upbeta^\sigma = \dfrac{\gamma}{w}.
\eea
Consistent with eq.~\eqref{eq:normal vector}, the definition of the lapse function requires $w(\sigma)>0$. Furthermore, the boost function $\gamma(\sigma)$ dictates the behaviour of the shift vector.

\subsubsection{The causal structure: ingoing and outgoing null vectors}\label{sec:nullvectors}
Outgoing and ingoing null vectors are given, respectively, by
\beq
\label{eq:conf_null_vectors}
\bar l^a = \nu \left( \delta^a_\tau - \dfrac{1+\gamma}{w}\delta^a_\sigma \right), \quad \bar k^a = \dfrac{w}{2\Xi \nu} \left( \delta^a_\tau + \dfrac{1 - \gamma}{w}\delta^a_\sigma \right),
\eeq
with $\nu(\sigma)$ a freely specifiable boost parameter under Lorentz transformation\footnote{The boost normalisation function $\nu(\sigma)$ in eq.~\eqref{eq:conf_null_vectors} should not be confused with the hyperboloidal boost function $\gamma(\sigma)$ from eq.~\eqref{eq:def_gamma}.
Even though, ref.~\cite{Zenginoglu:2011jz} originally introduced the terminology boost function for the quantity $\gamma(\sigma)$, the $3+1$ decomposition \eqref{eq:3+1} suggests that $\gamma(\sigma)$ could also be referred to as "shift function" to avoid confusion.}. 

To align either $\bar l^a$ or $\bar k^a$ with the vector $\left(\partial_\tau\right)^a$, one must choose the signs $-$ or $+$ in the conditions \eqref{eq:align_nullvectors_hrz}, respectvely. Indeed, conditions \eqref{eq:align_nullvectors_scri} imply the behaviour at future null infinity
\beq
\label{eq:nullvector_scri}
\left. \bar l^a\right|_{\sigma=0} = \nu \left( \delta^a_\tau- \dfrac{2}{w}\delta^a_\sigma \right), \quad \left. \bar k^a\right|_{\sigma=0}  =\dfrac{w}{2\Xi\nu} \delta^a_\tau,
\eeq
whereas at the black-hole horizon, the positive sign in eq.~\eqref{eq:align_nullvectors_hrz} provides
\beq
\left. \bar l^a\right|_{\sigma=\sigmah} = \nu \delta^a_\tau, \quad \left. \bar k^a\right|_{\sigma=\sigmah}  =\dfrac{w}{2\Xi\nu} \left( \delta^a_\tau  +\dfrac{2}{w}\delta^a_\sigma \right).
\eeq

\subsubsection{Trautman-Bondi Mass}
For asymptotically flat spacetimes, the Trautman-Bondi mass gives the energy content at future null infinity. We calculate this quantity via the asymptotic limit of the Hawking mass~\cite{Szabados:2009eka,Jaramillo:2010ay}. 

Given a closed $2-$surface ${\cal S}$ in the physical manifold ${\cal M}$, the Hawking Mass is defined as
\beq
M_{H} = \sqrt{\dfrac{{\cal A_{\cal S}}}{16 \pi} } \Bigg( 1 +  \dfrac{1}{8\pi} \oint_{{\cal S}}\Theta_+ \Theta_- dA  \Bigg),
\eeq
with ${\cal A_{\cal S}}$ the area of ${\cal S}$. Also, $\Theta_+$ and $\Theta_-$ are, respectively, the expansions of outgoing and ingoing null vectors given by
\beq
\Theta_{+} = \nabla_a l^a, \quad \Theta_{-} = \nabla_a k^a.
\eeq
The physical null vectors are obtained from the conformal re-scaling of eq.~\eqref{eq:conf_null_vectors} into  $l^a = \Omega \bar l ^a$ and $k^a= \Omega \bar k ^a$.

In the coordinates $(\tau,\sigma, \theta,\varphi)$, we parametrise the closed $2-$surface ${\cal S}$ by $\sigma =$ constant, $\tau$= constant, $(\theta, \varphi)\in [0,\pi]\times[0,2\pi]$. With the help of eqs.~\eqref{eq:eq_hyp} and \eqref{eq:conf_null_vectors}, the Hawking mass reads in terms of the hyperboloidal metric functions
\beq
M_{H} = \dfrac{\lambda}{2}  \Bigg[  \dfrac{\rho(\sigma) \Xi(\sigma)}{\sigma \beta(\sigma) } \bigg(1 - \dfrac{p(\sigma) \beta(\sigma)^2}{\Xi(\sigma) \sigma^2}  \bigg)\Bigg].
\eeq
The Trautman-Bondi mass follows by taking the limit to future null infinity along $\tau$= constant, i.e.
\bea
\label{eq:BondiMass}
M_{\rm TB} = \lim_{\sigma\rightarrow 0} M_H(\sigma).
\eea
With the definitions \eqref{eq:def_p} and \eqref{eq:def_Xi_0}, together with asymptotic expansions \eqref{eq:F_of_sigma_asymp_Mink} and \eqref{eq:zeta_asympot}, one verifies that the Trautman-Bondi mass coincides with the ADM mass defined via eq.~\eqref{eq:f_asymp_Mink}. While this result is expected due to the static character of the spacetime, eq.~\eqref{eq:BondiMass} provides a calculation of the mass direct from the hyperboloidal metric functions.

\subsubsection{The Schwarzschild spacetime: a first example}\label{sec:mingauge_schwarzschild}
To exemplify the above discussion we construct a hyperboloidal coordinate for the Schwarzschild spacetime. The line element \eqref{eq:metric_tr} has the functions $a (r)= b(r) = 1-\dfrac{\rh}{r}$ which already implies $\zeta(\sigma)=1$ in eq.~\eqref{eq:zeta}. Besides, eq.~\eqref{eq:def_tortoise} for the tortoise coordinate integrates to
\bea
\label{eq:r*_Schwarzschild}
r_*(r) &=& r + \rh \ln\left( \dfrac{r}{\rh} -1 \right) \nn \\
         &=& r + \rh \ln\left( \dfrac{r}{\rh}\right)  + \rh \ln\left( 1- \dfrac{\rh}{r} \right).
         \label{eq:r*_Schwarzschild}
\eea
The first line in the above expression corresponds to the well-known results in textbooks, whereas the second line makes explicit the behaviour of $r_*$ as $r\rightarrow \infty$ and $r\rightarrow \rh$.

To construct the height function $H(\sigma)$, we follow the strategy from refs.~\cite{Ansorg:2016ztf}. First, we introduce the ingoing Eddington-Finkelstein coordinate $v=t + r_*$. This step ensures the horizon penetrating character of the coordinate system. Along $v=$ constant, however, the limit $r\rightarrow\infty$ corresponds to past null infinity. The goal is to deform these coordinates, such that $r\rightarrow\infty$ reaches future null infinity. For this purpose we consider the asymptotic behaviour of outgoing null rays within the coordinate system $(v,r,\theta,\varphi)$. 

In coordinates $\{v,r\}$, the Lorentzian sector of eq.~\eqref{eq:metric_tr} becomes
\beq
\left. ds^2\right|_{{\cal M}^2} = dv \bigg( -a(r) dv^2 + 2 dr\bigg)
\eeq
with $\left.ds^2\right|_{{\cal M}^2} = 0$ characterising the null rays. By construction the surface $v=$constant $(dv=0)$ parametrises the ingoing light rays. Outgoing null rays satify
\bea
\label{eq:outgoing_rays_inv}
\dfrac{dv}{dr} &=& \dfrac{2}{1-\rh/r}  \nn \\
&= &2\left( 1 + \dfrac{\rh}{r} \right) + {\cal O}\left(\left(\dfrac{\rh}{r}\right)^2\right).
\eea
Considering just the leading order asymptotic behaviour, eq.~\eqref{eq:outgoing_rays_inv} integrates to
\beq
\label{eq:v_to_tau}
v = 2 r + 2 \rh \ln\left( \dfrac{r}{\rh} \right) + C.
\eeq
The hyperboloidal time coordinate $\tau$ is then chosen to be proportional to the integration constant $C=\lambda \tau$. Mapping eq.~\eqref{eq:v_to_tau} back into the original Schwarzschild time coordinate $t$ yields
\bea
t &=& v - r_* \nn \\
  &=& \lambda \tau + r +\rh\ln\left( \dfrac{r}{\rh}\right) - \rh\ln\left( 1-\dfrac{\rh}{r} \right).
  \label{eq:t_to_tau_in_r}
\eea
Before discussing the compactifcation function $r=r(\sigma),$ we are already able at this stage to identify the structure for a re-scaled height function $h:= \lambda H$ via $t = \lambda \tau - h(r)$. From eq.~\eqref{eq:t_to_tau_in_r}, it follows
\bea
\label{eq:h_r}
h(r) = -r  - \rh\ln\left( \dfrac{r}{\rh}\right) + \rh\ln\left(1- \dfrac{r}{\rh} \right).
\eea
The structure of eq.~\eqref{eq:h_r} is very similar to the tortoise coordinate $r_*(r)$ in eq.~\eqref{eq:r*_Schwarzschild}, with the only difference in the sign for the singular terms in the limit $r\rightarrow \infty$. This property is a direct consequence of retaining only the leading terms in eq.~\eqref{eq:outgoing_rays_inv}, as it will be demonstrated in sec.~\ref{sec:min_gauge}.

Finally, one must specify a compactification function $r(\sigma)$. The most simple strategy to parametrise the black-hole external region is
\beq
\label{eq:r_of_sigma}
r = \dfrac{\rh}{\sigma}.
\eeq
With this choice, the definitions \eqref{eq:HypCoord} and \eqref{eq:beta} yield the quantities
\beq
\label{eq:rho_beta_cte}
\rho(\sigma) = \beta(\sigma) = \dfrac{\rh}{\lambda}.
\eeq
 Besides, using eq.~\eqref{eq:r_of_sigma} in \eqref{eq:r*_Schwarzschild}, we read the dimensionless tortoise coordinate \eqref{eq:def_x}
 \beq
 \label{eq:tortoise_x_schwarzschild}
x(\sigma) = \dfrac{\rh}{\lambda}\left( \dfrac{1}{\sigma} -\ln\sigma + \ln(1-\sigma) \right).
 \eeq
 
Moreover, eqs.~\eqref{eq:t_to_tau_in_r} and \eqref{eq:r_of_sigma} provide the mapping $t=t(\tau,\sigma)$ from which one reads the height function 
\beq
\label{eq:Height_H}
H(\sigma) = \dfrac{\rh}{\lambda}\left( -\dfrac{1}{\sigma} + \ln \sigma + \ln\left( 1-\sigma \right) \right).
\eeq
It becomes evident once again, that the height function $H(\sigma)$ in the minimal gauge \eqref{eq:Height_H} follows from changing the sign of the singular components around $\sigma =0$ in the tortoise coordinate $x(\sigma)$ \eqref{eq:tortoise_x_schwarzschild}. This property is generic, as it will be showed in sec.~\ref{sec:min_gauge}

Even though this strategy was the one used in refs.~\cite{Schinkel:2013tka,Schinkel:2013zm,Ansorg:2016ztf,PanossoMacedo:2018hab,PanossoMacedo:2019npm}, we present here an alternative derivation of eq.~\eqref{eq:Height_H} based on a similar intuitive process. Now, instead of choosing the ingoing Eddington-Finkelstein coordinate $v=t + r_*$ as the first step, we opt to first consider the outgoing null coordinate $u=t - r_*$. This choice ensures that future null infinity is reached for $r \rightarrow \infty$. Along $u =$ constant, however, $r = \rh$ corresponds to the white-hole horizon. To deform this coordinate system, we proceed in a similar manner as before and consider the asymptotic behaviour of ingoing null rays within the coordinate system $(u,r,\theta,\varphi)$. 

In coordinates $\{u,r\}$, the Lorentzian sector of eq.~\eqref{eq:metric_tr} becomes
\beq
\left. ds^2\right|_{{\cal M}^2} = du \bigg( -a(r) du^2 - 2 dr\bigg)
\eeq
with $\left.ds^2\right|_{{\cal M}^2} = 0$ characterising the null rays. Since the surfaces $u=$constant $(du=0)$ parametrises the outgoing light rays, the ingoing null rays follows from
\bea
\label{eq:du_dsigma}
\dfrac{du}{dr} =- \dfrac{2}{1-\rh/r} \quad \stackrel{{\rm eq.~}\eqref{eq:r_of_sigma}}{\longrightarrow} \quad \dfrac{du}{d\sigma} = \dfrac{2 \rh}{\sigma^2 \left(1-\sigma\right)} = \dfrac{2\rh}{1-\sigma} +{\cal O}(1-\sigma). 
\eea
Note that the above expression make use of the compact radial coordinate \eqref{eq:r_of_sigma} to facilitate the expansion around the horizon.
Integrating the leading order behaviour of eq.~\eqref{eq:du_dsigma} around $\sigma=1$ yields
\beq
\label{eq:u_of_tau}
u = - 2\rh \ln\left(1-\sigma \right) + C.
\eeq
Once again, we can choose the the integration constant to be proportional to the new time coordinate $\tau$. Mapping eq.~\eqref{eq:u_of_tau} back to $t = u + r_*$ yields the height function $H(\sigma)$ as in eq.~\eqref{eq:Height_H}.

From eqs.~\eqref{eq:tortoise_x_schwarzschild} and \eqref{eq:Height_H} the metric components follow from the definitions \eqref{eq:def_p}, \eqref{eq:def_gamma} and \eqref{eq:def_w}
\beq
\label{eq:metric_func_mingauge_Schwarzschild}
p(\sigma) = \dfrac{\lambda}{\rh}\sigma^2(1-\sigma), \quad \gamma(\sigma) = 1-2\sigma^2, \quad w(\sigma) = \dfrac{4\rh}{\lambda}(1+\sigma).
\eeq
Historically, the particular construction of hyperboloidal coordinates presented in this section emerged from the ingenious intuition from M. Ansorg \footnote{Deceased Dez. 2016.}\cite{Schinkel:2013zm,PanossoMacedo:2014dnr}. Refs.\cite{PanossoMacedo:2018hab} then showed, that keeping only the leading order in eq.~\eqref{eq:du_dsigma} is a sufficient condition to align the null vectors \eqref{eq:conf_null_vectors} with the generators of the future null infinity and the event horizon. However, there is no guarantee {\em a priori} that dropping the higher order terms from \eqref{eq:outgoing_rays_inv} would necessarily lead to spacelike hypersurfaces of $\tau=$constant. Still, the properties discussed in eq.~\eqref{eq:prop_p}, \eqref{eq:spacelike}-\eqref{eq:align_nullvectors_hrz} are directly verified by eqs.~\eqref{eq:metric_func_mingauge_Schwarzschild}. In particular, one observes that $w(\sigma)>0$ even for $\sigma>1$, i.e., though the coordinate transformation was initially devised to parametrise the black-hole exterior region $\sigma \in[0,1]$, it allows us to extend the mapping inside the black hole up to the singularity $\sigma\rightarrow \infty$. This property will be further discussed and explored in sec.~\ref{sec:trumpet}. Besides, refs.~\cite{Ansorg:2016ztf,PanossoMacedo:2018hab,PanossoMacedo:2019npm} established the connection between this particular geometrical construction of hyperboloidal surfaces with the analytical approach introduced by Leaver~\cite{Leaver85,Leaver90} in the study of quasi-normal modes. 

\subsection{Hyperboloidal gauges}

The reasoning outlined the previous section provides the basics steps to construct hyperboloidal slices in the so-called minimal gauge. Sec.~\ref{sec:min_gauge} brings further generic considerations about this gauge. However, we point out that further gauges have also been employed in the literature. We  summarise some of these alternatives gauge here, with ref.~\cite{PanossoMacedo:2019npm} bringing further details about them. Alternative hyperboloidal coordinates will differ from the minimal gauge in either degrees of freedom: the choice for the spatial compactification function $r(
\sigma)$, and the height function $H(\sigma)$. For instance, the original scri fixing technique\cite{Zenginoglu:2007jw} employed  a compactification in which the conformal factor $\Omega$ encoded the radial degree of freedom, whereas we employ a compact radial coordinate $\sigma = \lambda \Omega$ directly adapted to the conformal factor.

The first intuitions to the construct height function $H$ were developed from hyperboloids in the Minkowski spacetime\cite{Moncrief2000}. A closer look to the geometrical properties of exact hyperboloids in Minkowski spacetime offers a geometrical approach to the construction of hyperboloidal surfaces in black-hole manifolds. In flat space, hyperboloids are characterised as hypersurfaces with constant mean curvature (CMC). Such a geometrical characterisation can be extended to a curved spacetime, with ref.~\cite{Malec:2009hg,Bizon:2008zd,Bizon:2010mp} employing CMC slices in the Schwarzschild spacetime and ref.~\cite{Cruz-Osorio:2010rsr} generalising the CMC foliation for any spherically symmetric, static spacetime. The CMC slices have also played an important role when constructing initial data on hyperboloidal hypersurfaces \cite{Andersson92,Andersson:1993we,Andersson:1994ng,Frauendiener:1998ud,Buchman:2009ew,Schinkel:2013zm}. However, the CMC foliation in a closed form is only available for spherically symmetric spacetimes, and one must resort to a numerical construction of CMC slices in the Kerr solution\cite{Schinkel:2013tka}. This limitation makes the CMC gauge less appealing for black hole perturbation theory. Alternatively, one can adapt the explicit coordinate transformation between the Minkowsky $(t,r)$ coordinates into exact hyperboloids. These strategy was mainly employed by R\`acz et al. in flat and  black-hole spacetimes, e.g. \cite{Fodor:2003yg,Fodor:2006ue,Racz:2011qu,Csukas:2021sia}. 

In ref.~\cite{Zenginoglu:2007jw}, Zenginoglu suggests several hyperboloidal gauges for the Schwarzschild and Kerr spacetime, including the CMC foliation mentioned above. The compactification function employed in his work satisfies the minimal gauge gauge condition for the radial coordinate, i.e. $\beta=$constant. However, the several choices for height functions discussed differs from the one in the minimal gauge, as detailed in ref.~\cite{PanossoMacedo:2019npm}. Moreover, the work sets the ground for the matching layers strategy, which allows the construction of hyperboloidal layers in the wave zone and/or black hole region that coincides with arbitrary spacelike foliations in the interior. This strategy was employed initially by refs. \cite{Bernuzzi:2010xj,Bernuzzi:2012ku} to enhance the prediction of GW wave forms with the effective-one-body approach. Despite the success of this strategy, the authors also noticed\cite{Harms:2014dqa} that matching layer strategy is not necessary in black-hole perturbation, and a single hyperboloidal slice extending all the way between future null infinity and the horizon is preferable. As demonstrated in ref.~\cite{PanossoMacedo:2019npm}, the gauge introduce in ref. \cite{Harms:2014dqa} differs from the minimal gauge just by an extra factor $\sim \sigma$ in the height function. All in all, alternative gauges (e.g. \cite{Andersson:2019dwi}) will differ from the minimal gauge height function $H^{\rm mg}(\sigma)$ by an overall regular function $A(\sigma)$, i.e. $H(\sigma) = H^{\rm mg}(\sigma) + A(\sigma)$.

The above discussion is restricted to asymptotically flat spacetimes. Ref.~\cite{Sarkar:2023rhp} introduces hyperboloidal coordinates to de Sitter spacetimes. Even thought the reasoning for construction the height function is similar to the one employed in the minimal gauge, the authors treatment for the radial coordinate differs from the one suggested in this framework at the conceptual level. Specifically, the authors opt for mapping the radial coordinate as
\beq
\label{eq:r_of_sigma_linear}
r(\sigma) = a + b\, \sigma,
\eeq 
which is nothing more than a simple re-scale of the interval $[r_h, r_c]$ into $[\sigma_h,\sigma_c]$. Their particular choice for the constants $a$ and $b$ ensures that the event horizon $r_h$ is mapped into $\sigma_h = 1$ and the cosmological horizon $r_c$ into $\sigma = 0$. If one wishes to represent eq.~\eqref{eq:r_of_sigma_linear} in terms of $\rho(\sigma)$ and $\beta(\sigma)$ as in eqs.~\eqref{eq:HypCoord}-\eqref{eq:beta}, it then follows that
\beq
\rho(\sigma) = \dfrac{\sigma}{\lambda}(a + b \sigma) \Longrightarrow  \beta(\sigma) = -\dfrac{b}{\lambda}\sigma^2,
\eeq
which is not in the minimal gauge. This choice is perfectly justified because the domain of interest is usually the exterior black-hole region $r\in[r_h, r_c]$. This approach, however, does not allow for a conformal representation of the spacetime in terms of the regular conformal metric \eqref{eq:conf_metric}. Indeed, the de Sitter boundary $r\rightarrow \infty$ is still at $\sigma\rightarrow \infty$.

\section{Black Hole Perturbation Theory}\label{sec:pert_theory}
To exemplify the effects of the hyperboloidal coordinates on the perturbation equations for black-hole perturbation theory, we consider scenarios in which the dynamics of the fields are reducible to a single master equation of the form
\beq
\label{eq:waveeq_Schwarzschild}
-\Psi_{,tt} + \Psi_{,r_* r_*} - {\cal V}(r)\,\Psi = R(r).
\eeq
The potential ${\cal V}(r)$ is assumed to vanish as $r_*\rightarrow \pm \infty$. We recall that the eq.~\eqref{eq:waveeq_Schwarzschild} is not necessarily the most generic expression to describe perturbations over spherically symmetric spacetimes, with line element of the form \eqref{eq:metric_tr} as discussed, for instance, in refs.~\cite{Liu:2022csl,Cardoso:2022whc}. However, eq.~\eqref{eq:waveeq_Schwarzschild} applies for a wide range of scenarios with spherical symmetry, and it provides us with an useful proxy for understanding how to apply the hyperboloidal methods in black-hole perturbation theory. 

The discussion in this section will focus on two independent approaches to tackle eq.~\eqref{eq:waveeq_Schwarzschild}: time and frequency domain calculations. We briefly review here these approached in the usual Schwarzschild coordinates $(t,r,\theta, \varphi)$, with focus on the boundary conditions for solutions to the homogeneous equation.

Since the ${\cal V}(r)$ vanishes asymptotically, homogenous solutions show the asymptotic behaviours $\Psi^{\rm hom.} \sim e^{-i \omega(t \pm r_*)}$. In particular, the retarded solution satisfies $\Psi^{\rm ret} \sim e^{-i \omega(t - r_*)}$ for $r_*\rightarrow \infty$ and $\Psi^{\rm ret} \sim e^{-i \omega(t + r_*)}$ for $r_*\rightarrow -\infty$. These boundary conditions ensure that energy flows into the horizon and out towards the wave zone, and they are precisely formulated in the differential form
\beq
\label{eq:BC_tr_time}
\lim_{r_* \rightarrow \pm \infty} \Bigg(  \dfrac{ \Psi^{\rm ret}_{,r_*}}{\Psi^{\rm ret}_{,t} } \Bigg) = \mp 1.
\eeq
In coordinates $(t,r, \theta, \varphi)$, the frequency domain equation follows from a Fourier Ansatz 
\beq
\label{eq:fourierTransf_tr_coord}
\Psi(t,r) = e^{-i \omega t} \psi(r)
\eeq
and also \beq
R(t,r) = e^{-i \omega t} {\cal R}(r).
\eeq
Applied to \eqref{eq:waveeq_Schwarzschild}, these transformations yield
\beq
\label{eq:fourierEq_tr_coord}
\psi_{,r_* r_*} - \left( {\cal V} - \omega^2 \right)\,\psi = {\cal R}.
\eeq
In the frequency domain, the Fourier decomposition \eqref{eq:fourierTransf_tr_coord} allows also to express the boundary conditions for the retarded field \eqref{eq:BC_tr_time} as
\beq
\label{eq:BC_tr_freq}
\lim_{r_* \rightarrow \pm \infty} \Bigg(  \dfrac{ \psi^{\rm ret}_{,r_*} }{\psi^{\rm ret}} \Bigg) = \pm i \omega.
\eeq

\subsection{Time domain}
To express the wave equation \eqref{eq:waveeq_Schwarzschild} in the hyperboloidal coordinates $(\tau,\sigma,\theta,\varphi)$, we first derive the following relation from the coordinate transformations \eqref{eq:HypCoord}
 and \eqref{eq:def_x}\footnote{We also express here the relation $\partial_r = -\dfrac{1}{\lambda {\cal F}(\sigma)} \bigg( \gamma(\sigma)\partial_\tau  + p(\sigma) \partial_\sigma  \bigg)$ for consistency and completeness. }

\bea
\label{eq:partial_der_transf}
\partial_t = \lambda^{-1} \partial_\tau, \quad 
\partial_{r_*} = - \lambda^{-1}  \bigg( \gamma(\sigma) \partial_\tau  + p(\sigma) \partial_\sigma \bigg).
\eea
Eq.~\eqref{eq:waveeq_Schwarzschild} then transforms to
\beq
\label{eq:AxialWaveEq_hyperboloidal}
-w\, \overline \Psi_{,\tau \tau} + \boldsymbol{L_1}  \overline \Psi + \boldsymbol{L_2} \overline \Psi_{,\tau}=  \overline R
\eeq
with 
\beq 
\label{eq:masterfunc_coordmapp}
\overline \Psi(\tau,\sigma) = \Psi(t,r)
\eeq 
and
\beq
\label{eq:rescaled_source}
\overline R(\tau, \sigma) = \dfrac{\lambda^2}{p(\sigma)} R(t,r).
\eeq
Besides, the operators $\boldsymbol{L_1}$ and $\boldsymbol{L_2}$ read
\bea
\label{eq:L1}
\boldsymbol{L_1} &=&   \partial_\sigma \bigg(  p(\sigma) \partial_\sigma  \bigg) - {\overline {\cal V}}(\sigma) ,\\
\boldsymbol{L_2} &=&   2\gamma(\sigma)\partial_\sigma + \gamma'(\sigma),
\eea 
with $w(\sigma)$, $p(\sigma)$ and $\gamma(\sigma)$ given by eqs.~\eqref{eq:def_p},\eqref{eq:def_gamma} and \eqref{eq:def_w}. The operator $\boldsymbol{L_1}$ assumes a Sturm–Liouville form, whereas $\boldsymbol{L_2}$ is responsible for the energy dissipation at the boundary~\cite{jaramillo2021pseudospectrum,Gasperin:2021kfv}.
The re-scaled potential reads 
\bea
\label{eq:HypFunc_Pot}
{\overline {\cal V}}(\sigma) &= \dfrac{\lambda^2}{p(\sigma)} {{\cal V}}(r).
\eea

With a first order reduction in time $\overline \Phi = \overline \Psi_{,\tau}$, one re-casts eq.~\eqref{eq:AxialWaveEq_hyperboloidal} in terms of an evolution operator $\boldsymbol{L}$ as
\begin{equation}
\label{matrix_eq_time_domain}
- \vec{U}_{,\tau} + \boldsymbol{L} \vec{U} = \vec{S},
\end{equation}
with
\begin{equation}\label{matrix evolution}
 	\boldsymbol {L} =
\left(
\begin{array}{c  c}
	{0} & {1} \\
	w^{-1}\boldsymbol{L}_1 & w^{-1}\boldsymbol{L}_2
\end{array}
\right), \quad \vec U=\left(
\begin{array}{c}\overline\Psi \\ \overline{\Phi} \end{array}\right), \quad
\vec S=\left(
\begin{array}{c}0 \\ w^{-1} \overline{R} \end{array}\right).
\end{equation}
Of particular relevance for the study of QNMs is the homogenous version of eq.~\eqref{matrix_eq_time_domain}
\begin{equation}
\label{eigenvalue_eq_time_domain}
 \vec{U}_{,\tau} = \boldsymbol{L} \vec{U}.
\end{equation}
To discuss the boundary condition in the hyperboloidal framework, we first use eq.~\eqref{eq:partial_der_transf} to transform the left-hand side of eq.~\eqref{eq:BC_tr_time} into
\bea
\label{eq:BC_tausigma_time}
\lim_{r_* \rightarrow \pm \infty} \bigg(\dfrac{ \Psi^{\rm ret}_{,r_*}}{\Psi^{\rm ret}_{, t}}\bigg) &=&   
 -  \gamma(\sigma_\mp) -  p(\sigma_\mp)  \left. \dfrac{ \bar \Psi^{\rm ret}_{,\sigma}   } {\bar \Psi^{\rm ret}_{,\tau} } \right|_{\sigma=\sigma_{\mp}}
%
%
%
\eea
The right-hand side of eq.~\eqref{eq:BC_tausigma_time} shows that we can evaluate the original boundary conditions \eqref{eq:partial_der_transf} {\em directly} at the null surfaces $\sigma_\mp$ corresponding to $r_* \rightarrow \pm \infty$. Most importantly, these conditions are trivially satisfied. Indeed, at these null surfaces $p(\sigma_\mp) = 0$, c.f. eq.~\eqref{eq:prop_p}, and conditions \eqref{eq:align_nullvectors_hrz}-\eqref{eq:align_nullvectors_scri} ensure that $\gamma(\sigma_\mp) = \pm 1$ so that the right-hand side of eq.~\eqref{eq:BC_tr_time} is straightfowardly obtained.  

The arguments above are only valid under the assumption that $\bar \Psi^{\rm ret}(\tau,\sigma)$ and its derivatives are finite at the boundaries. This requirement is a consequence of wave equation \eqref{eq:AxialWaveEq_hyperboloidal} having a different character in coordinates $(\tau,\sigma,\theta, \varphi)$ when compared to the original \eqref{eq:waveeq_Schwarzschild}. Because $p(\sigma) = 0$ at the null surfaces, the principal part of \eqref{eq:AxialWaveEq_hyperboloidal} degenerates at the boundaries, i.e., the term proportional to $\bar \Psi_{,\sigma \sigma}$ vanishes there. The hyperboloidal wave equation is singular, and  $\bar \Psi^{\rm ret}(\tau,\sigma)$ will correspond to the underlying regular solution. 
Following the  terminology from ref.~\cite{boyd2001chebyshev}, the resulting boundary conditions are behavioral instead of numerical in the hyperboloidal formulation. They follow by enforcing the equation's underlying regularity conditions at $\sigma=\sigma_\mp$. In practical terms, the boundary conditions for the retarded field $\bar \Psi^{\rm ret}(\tau,\sigma)$ result from simply imposing the homogenous wave equation in the entire domain $\sigma\in[\sigmai, \sigmaf]$. The argument then extends if one seeks the solution to the inhomogenous eq.~\eqref{matrix_eq_time_domain}. In this case, the regularity of the solution will be constraint by the regularity of the re-scaled source $\overline R(\tau, \sigma)$, c.f. eq.~\eqref{eq:rescaled_source}.

\subsection{Frequency domain}

The corresponding frequency domain equation within the hyperboloidal approach is obtained in two equivalent ways: (i) via direct Fourier transformation of the hyperboloidal wave equation eq.~\eqref{matrix_eq_time_domain}; or (ii) via a coordinate change and conformal re-scaling of the frequency domain eq.~\eqref{eq:fourierEq_tr_coord} in the original Schwarzschild-like coordinate.

The first approach follows from directly transforming eq.~\eqref{eq:AxialWaveEq_hyperboloidal} into the frequency domain via 
\beq
\label{eq:fourierTransf_tausigma_coord}
\overline \Psi(\tau,\sigma) = e^{s \tau} \overline \psi(\sigma),
\eeq
together with 
\beq
\overline  R(\tau, \sigma) = e^{s\tau} \overline {\cal R}(\sigma).
\eeq 
The dimensionless frequency $s$ relates to the Fourier frequency $\omega$ via 
\beq
\label{eq:def_s}
s=-i \lambda \omega.
\eeq The conformal field $\overline \psi(\sigma)$ then satisfies 
\beq\label{axial_eq}
\boldsymbol{A} \overline \psi = \overline {\cal R} ,
\eeq
with the operator $\boldsymbol{A}$
\beq
\label{eq:OperatorA_firstOrderRed}
\boldsymbol A = ( \boldsymbol{L_1} - w(\sigma) s^2) + s \boldsymbol{L_2}.
\eeq
We will present the expression for the source term $\overline{R}(\sigma)$ in the end of this section. For now, we concentrate on the homogenous equation $\boldsymbol{A} \overline \psi = 0$, for which eq.~\eqref{eigenvalue_eq_time_domain} yields the eigenvalue problem 
\begin{equation}\label{eigenvalue_eq}
	\boldsymbol{L} \vec{u} = s \vec{u}, \quad \vec u=\left(
\begin{array}{c}\overline\psi \\ \overline{\phi} \end{array}\right).
\end{equation}
Since this approach is applied directly to the conformal fields defined in terms of hyperboloidal coordinates, the boundary conditions for $\overline \psi(\sigma)$ are also behavioural. In other words, eq.~\eqref{axial_eq} is singular (principal part vanishes for $p(\sigma)=0$) and the physical scenario is described by the underlying regular solution.

The alternative option derives the hyperboloidal frequency domain equation directly from the equivalent ordinary eq.~\eqref{eq:fourierEq_tr_coord} in the Schwarzschild-like coordinate via a re-scaling 
\beq
\label{eq:FuncTransfHyp_FreqDomain}
\psi(r) = Z(\sigma) \, \overline \psi(\sigma).
\eeq

Finally, to map the operator from eq.~\eqref{eq:fourierEq_tr_coord} into \eqref{axial_eq} via the frequency domain re-scaling \eqref{eq:FuncTransfHyp_FreqDomain}, one needs the relations arising from the radial coordinate change in eq.~\eqref{eq:HypCoord} 
\bea
\dfrac{d}{dr_*} = - \dfrac{p(\sigma)}{\lambda} \dfrac{d}{d\sigma} \quad \Longleftrightarrow \quad \dfrac{d}{dr} = - \dfrac{\sigma^2}{\lambda \beta(\sigma)} \dfrac{d}{d\sigma}.
\eea
Even though this approach only refers to quantities defined on the radial coordinate, the function $Z(\sigma)$ inherits the geometrical construction from the hyperboloidal time transformation from eq.~\eqref{eq:HypCoord}. Considering the identity \eqref{eq:masterfunc_coordmapp}, one relates the Fourier Ans\"atze \eqref{eq:fourierTransf_tr_coord} and \eqref{eq:fourierTransf_tausigma_coord} via the coordinate mapping \eqref{eq:HypCoord} to obtain
\bea
e^{-i\omega t} \psi(r) &=& e^{-i\omega \lambda\left(\tau -H(\sigma) \right)} \psi(r)  \nn \\
&=& e^{s \tau}\underbrace{ e^{-s H(\sigma)} \psi(r)}_{\bar \psi(\sigma)}.
\eea
A direct comparison of the right-hand side in the above expression with the definition \eqref{eq:FuncTransfHyp_FreqDomain} yields
\beq
\label{eq:funcZ}
 Z(\sigma) = e^{s H(\sigma)}.
\eeq
With this result, one recovers the operator $\boldsymbol{A}$ from eq.~\eqref{eq:OperatorA_firstOrderRed}, which may be re-express into the alternative form
\bea
\boldsymbol A &=& \alpha_2 \dfrac{d^2}{d\sigma^2} +  \alpha_1 \dfrac{d}{d\sigma} +  \alpha_0, \label{eq:OperatorA_SecOrde}
\eea
with the coefficients
\bea
\alpha_2(\sigma) &=& p(\sigma), \label{eq:alpha_2_axial}\\
\alpha_1(\sigma) &=& p'(\sigma) + 2s\, \gamma(\sigma), \\
\alpha_0(\sigma) &=& - s^2\, w(\sigma) + s\, \gamma'(\sigma) - {\overline{\cal V}}(\sigma).
\eea
In addition to this, the source term reads
\beq
\label{eq:Source_Hyp}
\overline{\cal R}(\sigma) =  \dfrac{\lambda^2}{p(\sigma) Z(\sigma)} {\cal R}(r).
\eeq
While the form \eqref{eq:OperatorA_firstOrderRed} is more suitable for calculating QNMs~\cite{jaramillo2021pseudospectrum} through the eigenvalue system \eqref{eigenvalue_eq}, the representation \eqref{eq:OperatorA_SecOrde} is better adapted to the techniques introduced in ref.~\cite{PanossoMacedo:2022fdi} for solving the inhomogeneous equations with distributional and/or extended sources.

Recall that $\overline \psi(\sigma)$ is defined in the conformal manifold with metric $\overline g_{ab}$. As already mentioned, the boundary conditions for this function are derived directly by the equation's regularity conditions. In practical terms, the principal part of the operators degenerates at the boundaries with the vanishing of factor $\alpha_2(\sigma_\mp) = p(\sigma_\mp)$ and one can impose the ordinary differential equation \eqref{axial_eq} directly at $\sigma=\sigma_\mp$. 

The pre-factor $Z(\sigma)$ is responsible for taking into account the external boundary behaviours into the physical field $\psi(r)$ via eq.~\eqref{eq:FuncTransfHyp_FreqDomain}. Indeed, the frequency domain boundary condition \eqref{eq:BC_tr_freq} is mapped to
\bea
\lim_{r_* \rightarrow \pm \infty} \Bigg( \dfrac{  \psi^{\rm ret}_{,r_*} }{ \psi^{\rm ret} }\Bigg) &=&  - \dfrac{p(\sigma)}{\lambda} \Bigg( \dfrac{  Z'(\sigma_\mp)}{Z(\sigma_\mp)}  +  \dfrac{\bar \psi^{\rm ret}{}'(\sigma_\mp)  }{ \bar \psi^{\rm ret}(\sigma_\mp)} \Bigg) \nn \\
&=& -  \dfrac{s}{\lambda} \gamma(\sigma_\mp) - \dfrac{p(\sigma_\mp)}{\lambda} \dfrac{\bar \psi^{\rm ret}{}'(\sigma_\mp)  }{ \bar \psi^{\rm ret}(\sigma_\mp)} \label{eq:BC_freq_hyp}
\eea
The right-hand side of eq.~\eqref{eq:BC_freq_hyp} evaluates to the condition expressed in eq.~\eqref{eq:BC_tr_freq} thanks to eq.~\eqref{eq:prop_p}, conditions \eqref{eq:align_nullvectors_hrz}-\eqref{eq:align_nullvectors_scri} and the frequency re-scaling \eqref{eq:def_s}.

\section{The hyperboloidal Minimal Gauge}\label{sec:min_gauge}
In this section we lay out the procedure to construct hyperboloidal slices within a class referred to as the minimal gauge \cite{Schinkel:2013tka,Schinkel:2013zm,Ansorg:2016ztf,PanossoMacedo:2018hab,PanossoMacedo:2019npm}. It expands on the intuitive scheme presented in sec.~\ref{sec:mingauge_schwarzschild} and it consists in retaining the minimal structure for radial function $\rho(\sigma)$ and the height function $H(\sigma)$ needed to achieve a hyperboloidal foliation of the spacetime.

\subsection{The radial compactification degree of freedom}\label{sec:min_gauge_radial}
Motivated by eq.~\eqref{eq:rho_beta_cte}, we define the radial compactification in the minimal gauge by imposing $\beta(\sigma)$ to be constant~\cite{PanossoMacedo:2018hab}. From eq.~\eqref{eq:beta}, this choice yields
\beq
\label{eq:rho_mingauge}
\rho(\sigma) = \rho_0 + \rho_1 \sigma \Longrightarrow \beta(\sigma) = \rho_0.
\eeq
Without loss of generality, one can fix the black-hole horizon $\rh$ at the coordinate vaule $\sigmah=1$. Eqs.~\eqref{eq:HypCoord} and \eqref{eq:rho_mingauge} then yield
\beq
\label{eq:eventhorz_fix}
\rho(\sigmah)=\dfrac{\rh}{\lambda} \Longrightarrow \rho_0 = \dfrac{\rh}{\lambda} - \rho_1.
\eeq

The radial compactification simplifies even further if one opts to set $\rho_1=0$, i.e., with the mapping $r(\sigma)$ given by eq.~\eqref{eq:r_of_sigma}. 
However, one can exploit the freedom in the choice $\rho_1$ to fix relevant spacetime surfaces in the compact coordinate $\sigma$. For instance, refs.~\cite{PanossoMacedo:2018hab,PanossoMacedo:2019npm} show that one can choose $\rho_1$ such that the Cauchy horizon in Reisner-Nordstr\"on (or Kerr) spacetime is always fixed at a given coordinate value $\sigma_{c} \rightarrow \infty$, regardless of the charge (or spin) parameters. Here, we exploit this property to introduce trumpet-minimal gauge hyperboloidal hypersurface in sec.~\ref{sec:trumpet}.

The compactification along the radial direction is also expressed in terms of the dimensionless tortoise coordinate $x(\sigma)$ via eqs.~\eqref{eq:def_x} and \eqref{eq:dx_dsigma}. The specific form for $x(\sigma)$ depends on the spacetime under consideration, a behaviour determined by the particular form of ${\cal F}(\sigma)$. 

We recall that ${\cal F}(\sigma)=0$ at the horizons $\sigmahi$ and, for asymptotically flat spacetimes, at $\sigma=0$. Since eq.~\eqref{eq:dx_dsigma} behaves as $x' \sim 1/{\cal F}$, the roots of the function ${\cal F}$ will introduce singular terms to $x(\sigma)$. Let us express the dimensionless tortoise coordinate as 
\beq
\label{eq:x_reg+sing}
x(\sigma) = x_{\rm sing}(\sigma) + x_{\rm reg}(\sigma).
\eeq
In the next paragraphs, we discuss how to construct an explicit representation for the terms contributing to $x_{\rm sing}(\sigma)$. The regular piece will then follow from integrating
\bea
\label{eq:dxreg_dsigma}
x_{\rm reg}'(\sigma) &=& x'(\sigma) - x_{\rm sing}'(\sigma) \nn \\
			       &=& - \dfrac{\rho_0}{\sigma^2 {\mathcal F(\sigma)}} - x_{\rm sing}'(\sigma).
\eea

\paragraph{Asymptotically flat spacetimes:}
In asymptotically flat spacetimes, one term in $x_{\rm sing}(\sigma)$ comes from the behaviour of ${\cal F}$ around $\sigma=0$. Eq.~\eqref{eq:dx_dsigma} with the asymptotic behaviour \eqref{eq:F_of_sigma_asymp_Mink} lead to
\beq
x'(\sigma) = -\dfrac{\rho_0}{\sigma^2}\left( 1+ \dfrac{2M}{\lambda} \dfrac{\sigma}{\rho_0} + {\cal O}(\sigma^2)\right).
\eeq
Thus, the leading terms contribute with the singular quantities
\beq
\label{eq:x0}
x_0(\sigma) = \dfrac{\rho_0}{\sigma} - \dfrac{2M}{\lambda}\ln \sigma.
\eeq

\paragraph{Horizons:} Each horizon $\sigmahi$ contributes to $x_{\rm sing}(\sigma)$ with a respective $x_{{\rm h}_i}(\sigma)$ term. From the definition \eqref{eq:dx_dsigma} and the representation \eqref{eq:F_of_sigma_j} around a given horizon $\sigmahi$, one obtains
\beq
x'(\sigma) = -\dfrac{\rho_0}{\sigma^2 {\cal K}_i(\sigma) \left(1- \dfrac{\rhi}{r(\sigma)} \right) }.
\eeq
The integration around the horizon $\sigmahi$ yields
\beq
\label{eq:x_horizons}
x_{{\rm h}_i}(\sigma) =\dfrac{\rhi}{\lambda\, {\cal K}_{{\rm h}_i}} \ln\left| \sigma - \sigmahi \right|, \quad {\cal K}_{{\rm h}_i}={\cal K}_i(\sigmahi).
\eeq


\paragraph{Asymptotically de Sitter spacetimes:}
For asymptotically de Sitter spacetimes, the contribution around $\sigma=0$ to the dimensionless tortoise coordinate is actually regular. Indeed, around $\sigma=0$, eq.~\eqref{eq:dx_dsigma} with the asymptotic expansion \eqref{eq:F_of_sigma_asymp_dS} yields
\beq
\label{eq:x0_dS}
x'(\sigma) = \dfrac{3}{\Lambda \lambda^2 \rho_0}\left( 1 + {\cal O}(\sigma^2)\right) \Longrightarrow x_0(\sigma) = \dfrac{3}{\Lambda \lambda^2 \rho_0} \sigma,
\eeq
which is a term contributing only to the regular part $x_{\rm reg}(\sigma)$ in eq.~\eqref{eq:x_reg+sing}. All terms contributing $x_{\rm sing}(\sigma)$ come actually from the behaviours around the horizon as in eq.~\eqref{eq:x_horizons}. 

Given the importance of the cosmological horizon in asymptotically de Sitter spacetimes, we explicitly single out its contribution from eq.~\eqref{eq:x_horizons}. For that purpose, we label this horizon $\sigma=\sigma_{\Lambda}$ and we associate the index value $i=N_h$ in the product expansion showed in eqs.~\eqref{eq:func_f_horizons} or \eqref{eq:F_of_sigma}. The contribution to $x_{\rm sing}(\sigma)$ from $\sigma_{\Lambda}$ reads 
\beq
x_{\Lambda}(\sigma) =\dfrac{r_{\Lambda}}{\lambda\, {\cal K}_{\Lambda}} \ln\left| \sigma - \sigma_\Lambda \right|.
\eeq

All in all, the dimensionless tortoise coordinate reads

\beq
\label{eq:x_flat}
x(\sigma) = \left\{ 
	\begin{array}{cc}
	\displaystyle
	\sum_{i=0}^{N_{\rm h}} x_{h_i}(\sigma) + x_0(\sigma) + x_{\rm reg}(\sigma) & {\rm (Asymptotically flat)}, \\
	\\
	\displaystyle
	\sum_{i=0}^{N_{\rm h}-1} x_{h_i}(\sigma) + x_{\Lambda}(\sigma)  + x_{\rm reg}(\sigma) & {\rm (Asymptotically de Sitter)}.
	\end{array}
\right.
\eeq



\subsection{The height function}\label{sec:MinGauge_Height}
Following the presciption from sec.~\eqref{sec:mingauge_schwarzschild}, the height function in the minimal gauge $H(\sigma)$ arises from studying the behaviour of ingoing and outgoing null geodesics around the surfaces one wishes the hyperboloidal slice to intersect. 

For the Schwarzschild spacetime exemplified in sec.~\ref{sec:mingauge_schwarzschild}, this goal was achieved by either one of the procedures: (i) consider ingoing null geodesics $v = t + r_*$, and then integrate outgoing null geodesics asymptotically around future null infinity $\sigma=0$; or (ii) consider outgoing null geodesics $u = t - r_*$, and then integrate ingoing null geodesics around the black hole horizon $\sigma=1$. 

While in Schwarzschild, both approaches are equivalent, more generic spacetimes may display a richer structure, for instance with Cauchy and/or cosmological horizons. Thus, the strategy may need adaption to account for each one of the horizons in the spacetime. Here, we will discuss the approaches (i) and (ii) in a generic formalism, and then apply them to specific examples in sec.~\ref{sec:examples_spacetime}.

\subsubsection{The in-out strategy}
The in-out strategy is the approach originally described in refs.~\cite{Schinkel:2013tka,Schinkel:2013zm,Ansorg:2016ztf,PanossoMacedo:2018hab,PanossoMacedo:2019npm}. It consists of first considering {\em ingoing} null geodescis $v=t+r_*$, which ensures that the time hypersurface penetrates the black-hole horizon. If a Cauchy horizon also exists in the spacetime, such as in the Reisnner-Nordstr\"om case, this choice also ensures that the hypersurface intersects the Cauchy horizon. In the coordinate $(v,\sigma, \theta, \varphi)$, outgoing null geodesics follow from
\bea
\label{eq:out_null}
\dfrac{dv}{d\sigma}  =  -\dfrac{2 \lambda \rho_0}{\sigma^2{\cal F}(\sigma)}.
\eea
\paragraph{Asymptotically flat spacetimes\\}
We first consider asymptotically flat spacetimes, for which ${\cal F}(\sigma)$ behaves as eq.~\eqref{eq:F_of_sigma_asymp_Mink}. Eq.~\eqref{eq:out_null} expands around $\sigma=0$ as
\bea
\label{eq:out_null_asymptopic}
\dfrac{dv}{d\sigma}  =  -\dfrac{2 \lambda \rho_0}{\sigma^2}\Bigg( 1 + \dfrac{2M}{\lambda} \dfrac{\sigma}{\rho_0} + {\cal O}(\sigma^2)  \Bigg).
\eea
Considering only the leading terms, eq.~\eqref{eq:out_null_asymptopic} integrates to
\beq
v = \lambda \bigg( \tau  - {\cal H}_0(\sigma) \bigg), \quad {\cal H}_0(\sigma) = - \dfrac{2  \rho_0}{\sigma} + \dfrac{4M}{\lambda}\ln \sigma,
\eeq
where the integration constant is chosen to be the proportional to the new time coordinate $\tau$.
Mapping ingoing coordinate $v$ back to the Schwarzschild time $t=v-r_*$ yields
\beq
t = \lambda \bigg( \tau - {\cal H}_0(\sigma) - x(\sigma) \bigg) \Longrightarrow H(\sigma) = x(\sigma) + {\cal H}_0(\sigma).
\eeq
From eq.~\eqref{eq:x0}, one observes that ${\cal H}_0(\sigma) = - 2\,x_0(\sigma)$. Thus, as in the Schwarzschild case, the height function in the minimal gauge follows from the tortoise coordinate $x(\sigma)$ in eq.~\eqref{eq:x_flat} by just changing the sign from the singular contribution around $\sigma=0$ 
\beq
\label{eq:H_inout_flat}
H(\sigma) = \sum_{i=0}^{N_{\rm h}} x_{h_i}(\sigma) - x_0(\sigma) + x_{\rm reg}(\sigma).
\eeq

\paragraph{Asymptotically de Sitter spacetimes\\}
In asymptotically de Sitter spacetimes however, the contribution around $\sigma=0$ is regular, as discussed in sec.~\ref{sec:min_gauge_radial}. Thus, the outgoing null geodesics \eqref{eq:out_null} must be integrated in the vicinity of the Cosmological horizon. For that purpose, eq.~\eqref{eq:out_null} is reformulated with the help of eq.~\eqref{eq:F_of_sigma_j} as
\beq
\label{eq:out_null_asymptopic_dS}
\dfrac{dv}{d\sigma}  =  -\dfrac{2 \lambda \rho_0}{\sigma^2 {\cal K}_\Lambda(\sigma)}\left(1- \dfrac{r_\Lambda}{r(\sigma)} \right)^{-1}.
\eeq
Integration around $r_\Lambda=r(\sigma_\Lambda)$ provides
\beq
v = \lambda \bigg( \tau - {\cal H}_{\Lambda}(\sigma)\bigg), \quad {\cal H}_{\Lambda}(\sigma) = -2 \dfrac{r_{\Lambda}}{\lambda} \dfrac{1}{ {\cal K}_{\Lambda}}\ln\left|\sigma- \sigma_{\Lambda}\right|.
\eeq
As in the asymptotically flat case, we observe that ${\cal H}_{\Lambda}(\sigma) = -2 x_\Lambda(\sigma)$. Thus, the contribution of ${\cal H}_{\Lambda}(\sigma)$ when mapping the coordinate $v$ back into $t = v - r_*$ is just a sign change between the height function $H(\sigma)$ and the dimensionless tortoise coordinate $x(\sigma)$ in the singular part corresponding to the Cosmological horizon
\beq
\label{eq:H_inout_dS}
H(\sigma) = \sum_{i=0}^{N_{\rm h}-1} x_{h_i}(\sigma) - x_{\Lambda}(\sigma)  + x_{\rm reg}(\sigma).
\eeq

\subsubsection{The out-in strategy}
The out-in strategy is analogous to the procedure describe in the previous section, but the first step in the construction of height function is taken by considered {\em outgoing} null geodesics $u = t- r_*$. This choice ensure that that along $u=$ constant, the limit $r\rightarrow \infty$ leads to future infinity, regardless whether the spacetime is asymptotically flat, or de Sitter. For asymptotically de Sitter spacetimes, this option also ensures that the Cosmological horizon is intersected at $r=r_\Lambda$. On the other end of the surface $u=$ constant, event and Cauchy horizons are those associated with the white hole region. Thus, a hyperboloidal foliation will require modifications on the time coordinate around the horizons. 

Ingoing null geodesics in the $(u,\sigma,\theta, \varphi)$ coordinates is described by 
\bea
\dfrac{du}{d\sigma}  &=&  \dfrac{2 \lambda \rho_0}{\sigma^2{\cal F}(\sigma)} \nn \\
\label{eq:in_null}
&=& \dfrac{2 \rhi}{{\cal K}_{{\rm h}_i}}\left( \sigmahi - \sigma  \right)^{-1},
\eea
where we have employed representation eq.~\eqref{eq:F_of_sigma_j} in the second line to facilitate the study around a given horizon $\rhi=r(\sigmahi)$. Integrating \eqref{eq:in_null} around $\sigmahi$ and choosing the integration constant as proportional to the hyperboloidal time provide
\beq
\label{eq:height_u_hrz}
u = \lambda\bigg(\tau - {\cal H}_{{\rm h}_i}(\sigma)\bigg), \quad {\cal H}_{{\rm h}_i}(\sigma) =  2 \dfrac{\rhi}{\lambda \, {\cal K}_{{\rm h}_i}} \ln\left|\sigmahi -\sigma \right|.
\eeq
Mapping the outgoing coordinate $u$ back to the Schwarzschild time $t = u +r_*$ yields
\beq
t = \lambda \Bigg( \tau - \bigg({\cal H}_{{\rm h}_i}(\sigma) - x(\sigma)\bigg) \Bigg).
\eeq
The contribution to the height function from each horizon is ${\cal H}_{\rm i}(\sigma) = 2 x_{{\rm h}_i}(\sigma)$, c.f. eq.~\eqref{eq:x_horizons}. 

If procedure \eqref{eq:height_u_hrz} is applied to {\em all} horizons $\rhi$ 
 ($i=0\cdots N_{\rm h}$) in asymptotically flat spacetimes, then the height functions becomes
 \bea
 \label{eq:Houtin}
 H(\sigma) &=& \sum_{i=0}^{N_{\rm h}} {\cal H}_{{\rm h}_i}(\sigma) - x(\sigma) \nn \\
 		 &=& \sum_{i=0}^{N_{\rm h}} x_{h_i}(\sigma) - x_0(\sigma) - x_{\rm reg}(\sigma).
    \label{eq:H_outin_flat}
 \eea
As already discussed, in asymptotically de Sitter spacetimes the cosmological horizon $r_\Lambda$ is treated differently from black-hole horizons. Thus, applying procedure \eqref{eq:height_u_hrz} to the horizons $\rhi$ 
 ($i=0\cdots N_{\rm h}-1$) gives
  \beq
  \label{eq:H_outin_dS}
 H(\sigma) = \sum_{i=0}^{N_{\rm h}-1} x_{h_i}(\sigma) - x_\Lambda(\sigma) - x_{\rm reg}(\sigma).
 \eeq
 
 As for the in-out strategy, the height function in the minimal gauge displays just a change in sign from the singular contribution around $\sigma=0$ (asymptotically flat) or $\sigma=\sigma_{\Lambda}$ (asymptotically de Sitter) when compared against the dimensionless tortoise coordinate \eqref{eq:x_flat}. Eqs.~\eqref{eq:Houtin}-\eqref{eq:H_outin_dS}, however, also show the opposite sign in the regular term with respect to $x(\sigma)$.
 
 \subsubsection{In-out versus out-in}
 A direct comparison between the expressions for the height function in the in-out \eqref{eq:H_inout_flat} and the out-in \eqref{eq:H_outin_flat} strategies --- or equivalently \eqref{eq:H_inout_dS} and \eqref{eq:H_outin_dS} --- shows that they differ by a factor $2 x_{\rm reg}(\sigma)$. In the example of the Schwarzschild case, discussed in sec.~\ref{sec:mingauge_schwarzschild}, $x_{\rm reg}(\sigma) = 0$ and both strategies provide the same results. All in all, these strategies will be equivalent whenever  $x_{\rm reg}(\sigma)$ is an overall constant. Such a constant only shifts the origin of the time coordinate $\tau$, and therefore, it does not contribute to the geometrical properties of the hyperboloidal slices. Indeed, eq.~\eqref{eq:def_gamma} shows that the boost function $\gamma(\sigma)$ depends only on derivatives of $H(\sigma)$. 
 
 For more intricate solutions, the contribution of a non constant regular part $x_{\rm reg}(\sigma)$ may change some geometrical properties of the hyperboloidal hypersurfaces.  Indeed, from the identity 
 \beq
 \label{eq:Hinout-Houtin}
 H^{\io} -H^{\oi} = 2 \, x_{\rm reg},
 \eeq
 one derives
 \bea
 \label{eq:gamma_inout_outin}
 \gamma^{\io} - \gamma^{\oi} &=& 2 \, x_{\rm reg}' \, p, \\ 
  \label{eq:w_inout_outin}
 w^{\io}- w^{\oi} &=& - 2 x_{\rm reg}' ( \gamma^{\io} +  \gamma^{\oi}).
 \eea
 Thus, in some cases one approach will be more favourable towards the other, specifically when leading to the violation of conditions \eqref{eq:spacelike}, \eqref{eq:align_nullvectors_hrz} and \eqref{eq:align_nullvectors_hrz}.

\section{Example of spacetimes}\label{sec:examples_spacetime}
We now expand on the example from section \ref{sec:mingauge_schwarzschild}, where the hyperboloidal minimal gauge for Schwarzschild was introduced. We apply the generic formalism developed for for the minimal gauge to several spacetimes to exemplify the several features discussed. First, we revisit the minimal gauge on Schwarzschild and discuss the role of the parameter $\rho_1$ in compactification function eq.~\eqref{eq:rho_mingauge}. The second case discusses the case of high-dimensional Schwarzschild black-hole, where the in-out and out-in strategies lead to two alternative hyperboloidal foliations in the minimal gauge. Then, the solution modelling a central black hole with a dark matter halo from ref.~\cite{Cardoso:2021wlq} brings an example in which the minimal gauge strategy only works with out-in strategy. Finally, we study the Reisnner-Nordstrom-de Sitter spacetime to exploit the role played by several horizons in the construction of the hyperboloidal minimal gauge.

\subsection{Singularity-approaching hyperboloidal slices in Schwarzschild}\label{sec:trumpet}
Section \ref{sec:mingauge_schwarzschild} constructed the minimal gauge hyperboloidal slice for Schwarzschild. The original goal focused on parametrising the black-hole exterior region, defined in the radial domain $\sigma\in[0, 1]$. One observers, however, that the minimal gauges allows us to analytically extend the time foliation into the interior region $\sigma \in [0, \infty)$, with $\sigma \rightarrow \infty$ locating the singularity. Indeed, it is clear from eq.~\eqref{eq:metric_func_mingauge_Schwarzschild} that $w(\sigma)>0$ for $\sigma\in[0,\infty)$, so that the surfaces $\tau=$ constant remain spacelike all the way from future null infinity up to the singularity. 

\begin{figure}[b!]
	\centering
	\includegraphics[scale=0.7]{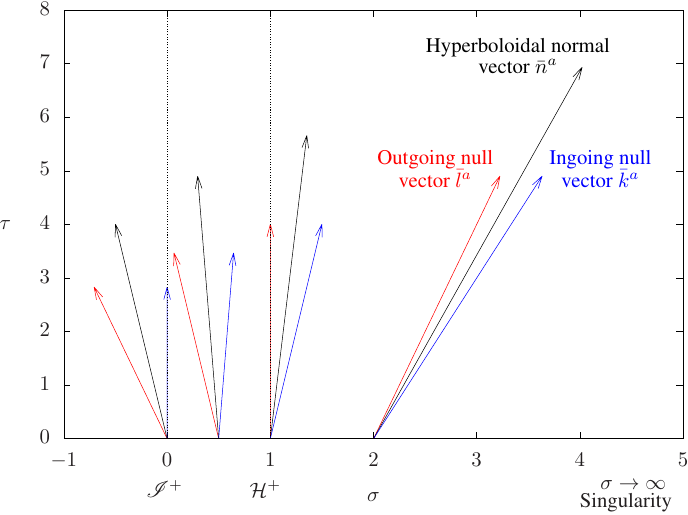}
	\caption{Causal structure for the Schwarzschild space time in the hyperboloidal minimal gauge $(\tau,\sigma)$-plane. Outgoing null vector $\bar k^a$ (blue) aligns with with the generator of time translation $\partial_\tau$ at future null infinity ($\sigma=0$). At the horizon $\sigma=1$,  the ingoing null vectors $\bar l^a$ (red) becomes parallel to $\partial_\tau$. At these boundaries, the light-cones point outwards the black-hole exterior region so that no boundary information propagates into the domain $\sigma\in[0,1]$. Besides, one can analytical extend the radial domain up to the singularity $\sigma\rightarrow \infty$. Indeed, the hyperboloidal normal vector $\bar n^a$ (black) remains timeline in the entire region $\sigma\in[0,\infty)$. Null vectors normalised to $\bar l^\tau = \bar k^\tau$, normal vector $\bar n^a$ normalised to unit $|| \bar n^a ||=-1$, and length parameter fixed to $\lambda = 2 \rh$.}
	\label{fig:SchwarzschildVectors}
\end{figure}

With the help of the null vectors $\bar l^a$ and $\bar k^a$ from eq. \eqref{eq:conf_null_vectors}, together with the hyperboloidal normal vector $\bar n^a$ from eq. \eqref{eq:normal vector},  Fig.~\ref{fig:SchwarzschildVectors} illustrates this propriety by showing the spacetime causal structure in the $(\tau,\sigma)$-plane. it becomes evident the alignment of the outgoing $\bar k^a$ and ingoing $\bar l^a$ null vectors with the time direction, respectively, at $\sigma=0$ and $\sigma=1$. At the boundaries $\sigma=0$ and $\sigma=1$, the light-cones point outwards the domain $\sigma\in[0,1]$, so that no information from the boundaries propagates into the black-hole exterior region. Besides, one observers that $\bar n^a$ remains inside the light cone formed out of $\bar l^a$ and $\bar k^a$ within the whole domain $\sigma\in[0,\infty)$. 

In this section we show that the minimal gauge is flexible to fix the singularity at a finite coordinate distance $\sigma_0$. Indeed, eq.~\eqref{eq:rho_mingauge} shows that there are two free parameters, namely $\rho_0$ and $\rho_1$, when choosing the radial compactfication $r(\sigma)$. The former is fixed by imposing that the black-hole horizon $\rh$ is at the coordinate $\sigma=1$, c.f.~eq~.\eqref{eq:eventhorz_fix}. In sec.~\ref{sec:mingauge_schwarzschild}, the later is fixed to $\rho_1=0$, and the physical singularity $r=0$ is mapped into $\sigma_o \rightarrow \infty$. 
Here we follow the strategy devised in refs.~\cite{PanossoMacedo:2018hab,PanossoMacedo:2019npm} and exploit the extra degree of freedom to extend the hyperboloidal foliation up to the singularity, by mapping the surface $r=0$ into the coordinate value $\sigma=\sigma_o>1$. This property follows for
\beq
\rho_1 = \dfrac{\rh}{\lambda} \dfrac{1}{1-\sigma_o}.
\eeq
Thus, the complete radial transformation \eqref{eq:HypCoord} with $\rho(\sigma)$ in the minimal gauge \eqref{eq:rho_mingauge} is given by
\beq
 r = \dfrac{\rh}{\sigma}\dfrac{1-\sigma/\sigma_o}{1 - 1/\sigma_o}.
\eeq
With this choice, the dimensionless tortoise coordinate $x(\sigma)$ mapped directly from $r_*(r)$ in eq.~\eqref{eq:r*_Schwarzschild} reads
\bea
\label{eq:x_trumpet}
&x(\sigma) = \dfrac{\rh}{\lambda}\Bigg[ \dfrac{1-\sigma/\sigma_o}{\sigma(1 - 1/\sigma_o)} + \ln\left( \dfrac{1-\sigma}{\sigma(1-1/\sigma_o)}  \right) \Bigg]  \\
	       &= \dfrac{\rh}{\lambda}\Bigg[ \dfrac{1}{\sigma(1 - 1/\sigma_o)}  - \ln  \sigma   + \ln\left| 1-\sigma \right|   - \left(  \dfrac{1}{\sigma_o(1 - 1/\sigma_o)} + \ln\left( 1-1/\sigma_o  \right)    \right)  \Bigg].\nn       
\eea
From the second line in the expression above we identify the contributions coming from future null infinity as in eq.~\eqref{eq:x0}
\beq
x_0(\sigma) =\dfrac{\rh}{\lambda}\Bigg( \dfrac{1}{\sigma(1 - 1/\sigma_o)}  - \ln  \sigma\Bigg),
\eeq
and, as expressed in eq.~\eqref{eq:x_horizons}, the contribution from horizon
\beq
x_{\rm h}(\sigma)= \dfrac{\rh}{\lambda} \ln\left| 1-\sigma \right|.
\eeq
With these two individual terms, one observes that $x'_{\rm reg} = 0$ in eq.~\eqref{eq:dxreg_dsigma}. Indeed, the regular piece in eq.~\eqref{eq:x_trumpet} is just the constant 
\beq
x_{\rm reg}(\sigma)=- \dfrac{\rh}{\lambda}\ \left(  \dfrac{1}{\sigma_o(1 - 1/\sigma_o)} + \ln\left( 1-1/\sigma_o  \right)    \right),
\eeq 
which ensures that $x\rightarrow 0$ as $\sigma\rightarrow \sigma_o$.

The hyperboloidal foliation is now valid in the domain $\sigma\in[0,\sigma_o]$, with $\sigma=0$ corresponding to $\scri^+$, $\sigma=1$ the black-hole horizon, and $\sigma=\sigma_o>1$ the singularity. With this choice the functions in eqs.~\eqref{eq:def_p}, \eqref{eq:def_gamma} and \eqref{eq:def_w} become
\bea
\label{eq:metric_funcs_trumpet_Schwarzschild}
p(\sigma)&=&\dfrac{\lambda}{\rh}\dfrac{\sigma^2 (1-\sigma) (1-1/\sigma_o)}{1-\sigma/\sigma_o}, \nn \\
\upgamma(\sigma) &=& \dfrac{(1-2\sigma^2) - (1 - 2\sigma)\sigma/\sigma_o}{1 - \sigma/\sigma_o}, \\
w(\sigma) &=& \dfrac{4 \rh}{\lambda} \dfrac{1+\sigma(1 - 1/\sigma_o)}{1-\sigma/\sigma_o}. \nn
\eea
In the limit $\sigma_o\rightarrow \infty$, the above expressions reduce to the ones in refs.~\cite{Ansorg:2016ztf,PanossoMacedo:2018hab,PanossoMacedo:2019npm}, c.f. eq.~\eqref{eq:metric_func_mingauge_Schwarzschild}.

\begin{figure}[h!]
	\centering
	\includegraphics[scale=0.7]{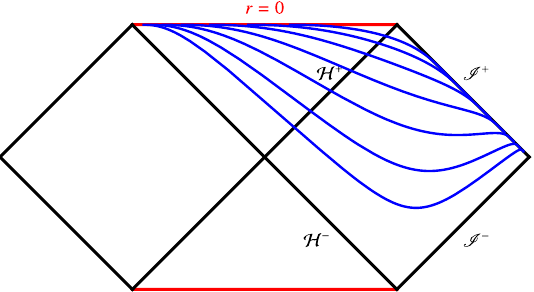}
	\includegraphics[scale=0.5]{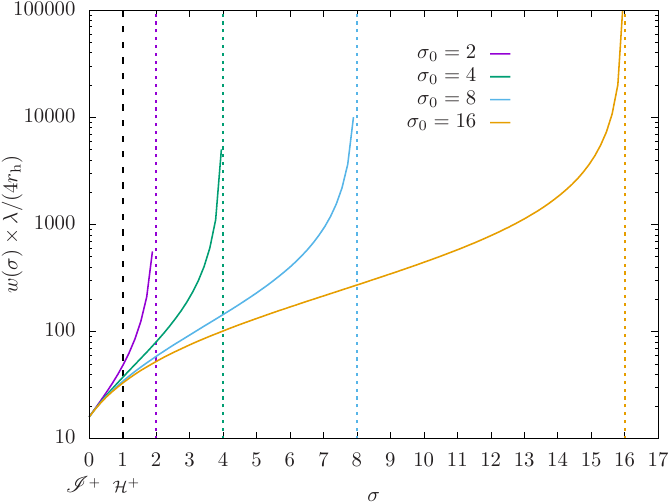}	
	\caption{{\em Left Panel:} Carter-Penrose diagram for the Schwarzschild spacetime with the hyperboloidal foliation (blue lines) extending between the singularity at $\sigma_o=2$, and $\scri^+$ at $\sigma=0$. The lapse behaves as $\alpha \sim \sqrt{\sigma_o - \sigma}$, the typical behaviour of trumpet slices. {\em Right Panel:} Metric function $w(\sigma)$ for different coordinate values $\sigma_0$ locating the physical singularity, marked by the colored dotted lines lines accordingly. The event horizon is fixed at $\sigma=1$ (black dashed line). The behaviour $w(\sigma)>0$ for $\sigma \in [0,\infty)$ ensure that the hyperboloidal time surfaces $\tau=$constant are spacelike in the entire domain, with metric components becoming singular at the physical singularity.}
	\label{fig:trumpet}. 
\end{figure}

Of particular interest is the behaviour of the lapse function $\alpha \sim \sqrt{\sigma_o - \sigma}$, c.f. \eqref{eq:3+1}, which is typical for a so-called trumpet slice \cite{Dennison:2014sma}. Even though such slices may play an important role when performing full non-linear evolutions of the Einstein's equations in the hyperboloidal setup, a complete study of this foliation goes beyond the scope of this work (see ref.~\cite{Vano-Vinuales:2023yzs}). The left panel of figure \ref{fig:trumpet} displays the Penrose diagram for the Schwarzschild spacetime with the compact hyperboloidal coordinates extending between the singularity at $\sigma=\sigma_o$ and future null infinity at $\sigma=0$. The right panel of figure \ref{fig:trumpet} shows the metric function $w(\sigma)$ for different choices of the singularity location $\sigma_0$. One observes $w(\sigma)>0$ in the entire domain $\sigma\in[0,\sigma_0)$, ensuring that the hyperboloidal hypersurfaces remain space-like all the way from future null infinity to the physical singularity.

\subsection{Higher dimension spacetimes: the Schwarzschild-Tangherlini spacetime}{\label{sec:HigDimSch}}
The Schwarzschild solution in spacetimes dimensions $d$ higher than $4$ provides an useful exercise to expand on the minimal gauge calculation laid out in sec.~\ref{sec:min_gauge}. In particular, it is the first example in which the in-out and out-in strategies give different results.

The line element for $d>4$ reads
\beq
\label{eq:metric_tr_highdim}
ds^2 = - f(r) dt^2 + \dfrac{dr^2}{f(r)} + r^2 d\omega^2_{d-2}, \quad f(r) = 1 -\left(\dfrac{\rh}{r}\right)^{d-3}
\eeq
with $d\omega^2_{d-2}$ the metric for the unit sphere in $d-2$ dimensions \cite{Tangherlini:1963bw}. We work in the minimal gauge with the compactficaiton function \eqref{eq:rho_mingauge}. Then, eq.~\eqref{eq:F_of_sigma} assumes the form
\bea
{\cal F}(\sigma) &=& (1-\sigma^{d-3}) \nn \\
\label{eq:F_highdim}
		 	&=&	  (1-\sigma)\, {\cal K}(\sigma), \quad {\cal K}(\sigma) = \sum_{k=0}^{d-4} \sigma^k.
\eea
To construct the dimensionless tortoise coordinate $x(\sigma)$, one needs to modify slightly the contribution from $\sigma\rightarrow 0$ originally given by eq.~\eqref{eq:x0}. Indeed, expanding eq.~\eqref{eq:dx_dsigma} around $\sigma=0$ leads to
\beq
x'(\sigma) = -\dfrac{\rh}{\lambda} \Bigg( \sigma^{-2} + (d-3) \sigma^{d-5} + {\cal O}(\sigma^{d-4}) \Bigg).
\eeq
When $d>4$, the singular contribution comes only from the leading term $\sigma^{-2}$, and therefore
\beq
x_0(\sigma) = \dfrac{\rh}{\lambda\, \sigma},
\eeq
i.e. the logarithmic term from eq.~\eqref{eq:x0} is present only for $d=4$. The singular contribution to $x(\sigma)$ arising from the horizon, though, follows directly from eq.~\eqref{eq:x_horizons}
\beq
x_{\rm h}(\sigma) = \dfrac{\rh}{\lambda}\dfrac{\ln|1-\sigma|}{d-3}.
\eeq
After removing the singular contributions as in eq.~\eqref{eq:dxreg_dsigma}, one observes that the regular term $x_{\rm reg}(\sigma)$ is non-trivial and defined by the differential equation
\beq
\label{eq:dx_reg_dsigma_highdim}
x'_{\rm reg}(\sigma) = \dfrac{\rh}{\lambda} \Bigg(  \dfrac{1}{d-3} - \dfrac{\sigma^{d-5}}{\cal K(\sigma)}\Bigg)  \left(1-\sigma\right)^{-1}.
\eeq
From eq.~\eqref{eq:F_highdim} it follows ${\cal K}(0)=1$ and ${\cal K}(1)=d-3$. These properties ensure that eq.~\eqref{eq:dx_reg_dsigma_highdim} is indeed regular in the interval $\sigma\in[0,1]$. 

Due to the non-trivial character of the regular term $x_{\rm reg}(\sigma)$, the height function in the minimal gauge differs in the in-out and out-in strategies. With the singular contributions $x_0(\sigma)$ and $x_{\rm h}(\sigma)$ the height functions read from eqs.~\eqref{eq:H_inout_flat} and \eqref{eq:H_outin_flat}, respectively
\beq
\label{eq:H_MinGague_IO_OI}
H^{\io}(\sigma) = - x_0(\sigma) + x_{\rm h}(\sigma) + x_{\rm reg}(\sigma), \quad H^{\oi}(\sigma) = - x_0(\sigma) + x_{\rm h}(\sigma) - x_{\rm reg}(\sigma)
\eeq
The explicit form of $x_{\rm reg}(\sigma)$ is not needed to construct the line element functions \eqref{eq:def_p}, \eqref{eq:def_gamma} and \eqref{eq:def_w}. The function $p(\sigma)$ is independent of the height function, and it reads
\beq
p(\sigma)= \dfrac{\lambda}{\rh}\sigma^2\left( 1-\sigma^{d-3} \right).
\eeq
For the in-out strategy, the functions $\gamma(\sigma)$ and $w(\sigma)$ assume the rather simple form
\beq
\gamma^{\io}(\sigma) = 1 - 2 \sigma^{d-3}, \quad w^{\io}(\sigma) = \dfrac{4\rh}{\lambda}\sigma^{d-5}.
\eeq
Conditions \eqref{eq:spacelike}, \eqref{eq:align_nullvectors_hrz} and \eqref{eq:align_nullvectors_scri} are all satisfied since $\gamma^{\io}(\sigma)$ has a similar structure to its $d=4$ counterpart in eq.~\eqref{eq:metric_func_mingauge_Schwarzschild}. However, the expression in the limit $d\rightarrow 4$ is not directly recovered because the minimal gauge includes only the leading order term in $x_0(\sigma)$. As explained, $x_0(\sigma)$ contains a logarithmic divergence in $d=4$, which is absent when $d>4$. 

\begin{figure}[t!]
	\centering
	\includegraphics[scale=0.57]{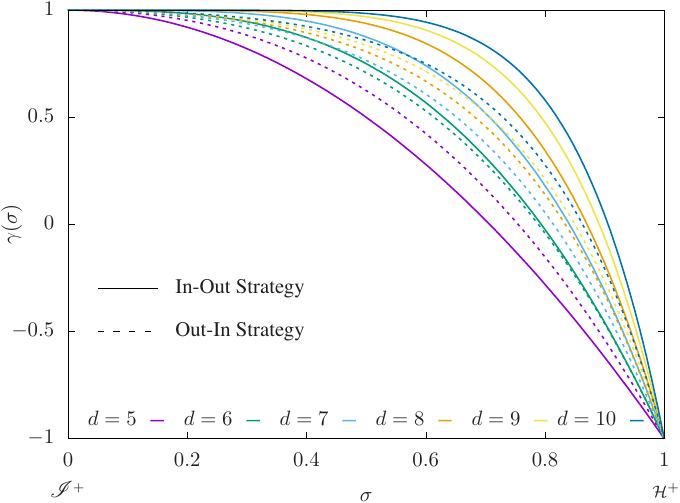}
	\includegraphics[scale=0.57]{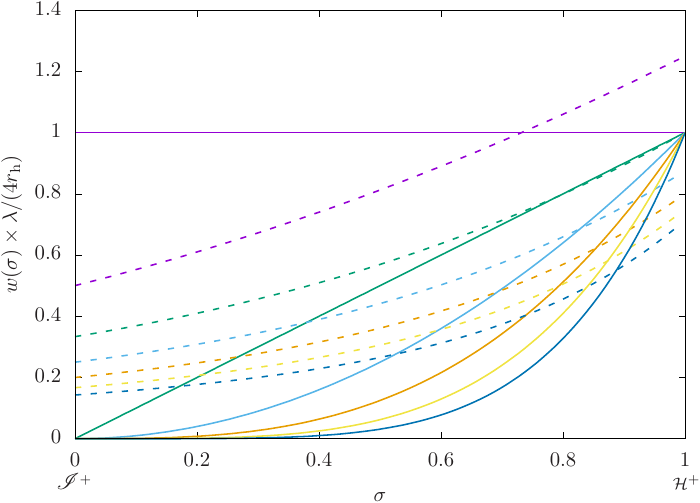}
	\caption{Metric functions $\gamma(\sigma)$ and $w(\gamma)$ for higher-dimensional Schwarzschild black holes in hyperboloidal slices. Contrary to the $4$-dimensional Schwarzschild solution, the in-out (solid lines) and out-int (dashed lines) strategies to construct the minimal gauge provide different results when $d>4$. For both choices conditions \eqref{eq:spacelike}, \eqref{eq:align_nullvectors_hrz} and \eqref{eq:align_nullvectors_scri} are satisfied. Moreover, $w(0)=0$ for $d>5$ with the in-out strategy, indicating that these slices are asymptotically null.}
	\label{fig:gamma_w_highdim}
\end{figure}

It is also intersting to observe that the function $w^{\io}(\sigma)$ vanishes at $\sigma=0$ for $d>5$. This property indicates that the hypersurfaces with $\tau =$ constant becomes null at future null infinity when $d>5$. Recall that eqs.~\eqref{eq:nullvector_scri} provide the expression for the outgoing and ingoing conformal null vectors at $\scri^+$. To ensure a regular expression for their components, one must choose the parameter $\nu(\sigma) = w(\sigma)/2$, which leads to 
\beq
\label{eq:nullvector_scri_highdim}
\left. \bar l^a\right|_{\sigma=0} = - \delta^a_\sigma, \quad \left. \bar k^a\right|_{\sigma=0}  = \delta^a_\tau \quad (d>5).
\eeq
Expressions \eqref{eq:nullvector_scri_highdim} confirms that, at future null infinity, not only does the ingoing null vector $\bar k^a$ aligns with the coordinate basis vector $\partial_\tau$, but also the outgoing null vector $\bar l^a$ aligns with the $\partial_\sigma$.

These properties change if one follows the out-in strategy, from which one derives
\beq
\gamma^{\oi}(\sigma) = 1  - 2   \dfrac{\sigma^2{\cal K}(\sigma)}{d-3}, \quad
 w^{\oi}(\sigma) = \dfrac{4\rh}{\lambda (d-3)} \Bigg(1 - \sigma^2 \dfrac{{\cal K}(\sigma)}{d-3} \Bigg)\left( 1-\sigma\right)^{-1}.
\eeq
Though less straightforward, one can again verify that conditions \eqref{eq:spacelike}, \eqref{eq:align_nullvectors_hrz} and \eqref{eq:align_nullvectors_scri} are all satisfied. With this strategy, however,  $w^{\oi}(0)$ never vanishes and the hyperboloidal surfaces are spacelike in the entire domain $\sigma\in[0,1]$ for all dimensions $d$. Fig.~\ref{fig:gamma_w_highdim} shows the functions $\gamma(\sigma)$ and $w(\sigma)$ for the in-out (solid lines) and out-in (dashed lines) strategies for $d=5\cdots 10$.

\subsection{Black-hole + matter halo}\label{sec:BHHalo}
Ref.~\cite{Cardoso:2021wlq} introduced a family of solutions to Einstein's equations with gravity minimally coupled to an anisotropic fluid. These asymptotically flat spacetimes describe a regular horizon with "hair", and they model the geometry of supermassive black holes with a dark matter halo. The functions in the line element ~\eqref{eq:metric_tr} read
\beq
a(r) = \left( 1- \dfrac{2 M_{\rm BH}}{r} \right) e^{\Upsilon(r)}, \quad b(r) = 1 - \dfrac{2 m(r)}{r},
\eeq
with the red-shift $\Upsilon(r)$ and mass functions $m(r)$ given by
\bea
\Upsilon(r)&=&\sqrt{\frac{M}{\xi}}\Bigg(2\arctan\left(\frac{r+a_0-M_0}{\sqrt{M_0\xi}}\right)-\pi\Bigg), \\ 
m(r) &=& M_{\rm BH} + \dfrac{M_0 r^2}{(a_0 +r)^2}\left(1 - \dfrac{2M_{\rm BH}}{r} \right)^2.
\eea
The parameters of the solution are the black-hole mass $M_{\rm BH}$, the mass $M_0$ of a "halo" surrounding the black hole, and a typical lengthscale $a_0$ for the matter distribution. Astrophysical scenarios are restricted to the cases with $\xi=2 a_0 - M_0 + 4 M_{\rm BH} > 0$. The black-hole radius reads $\rh=2M_{\rm BH}$. 

This is the first example in which the functions $a(r)$ and $b(r)$ in eq.~\eqref{eq:metric_tr} do not coincide. To employ the formalism developed in the previous sections, it is more convenient to re-express
\beq
b(r) = \left( 1- \dfrac{r_{\rm h}}{r} \right) B(r), \quad B(r)= 1 - 2\left( 1- \dfrac{r_{\rm h}}{r}\right)\dfrac{M r}{(a_0+r)^2},
\eeq
from which the function $f(r)$ defined in eq.~\eqref{eq:def_tortoise} assumes directly the form \eqref{eq:func_f_horizons}
\beq
\label{eq:f_BHHalo}
f(r) = \left( 1- \dfrac{r_{\rm h}}{r} \right) K(r), \quad K(r) = \sqrt{e^{\Upsilon(r)} B(r)}.
\eeq
Comparing the asymptotic expansion of eq.~\eqref{eq:f_BHHalo} against eq.~\eqref{eq:f_asymp_Mink}, one can read off the total ADM mass $M = M_{\rm BH} + M_0$. As expected, the same result follows if one uses eq.~\eqref{eq:BondiMass} to calculate the Trautman-Bondi mass.
Following the parametrisation employed in refs.~\cite{Cardoso:2021wlq,Cardoso:2022whc}, we re-scale all dimensional quantities in terms of the black-hole mass $M_{\rm BH}$, so that $M_0 = \mu\, M_{\rm BH},$ $a_0 = \upalpha_0 \, M_{\rm BH}$ and $\xi = \upxi \, M_{\rm BH}$.

To construct a hyperboloidal foliation in the minimal gauge, we first compactify the radial coordinate according to eq.~\eqref{eq:r_of_sigma}. In terms of the compact radial coordinate, the relevant metric functions read
\bea
\upupsilon(\sigma) &=& \Upsilon(r(\sigma)) \nn\\
			     &=& \sqrt{\frac{\mu}{\upxi}}\Bigg(2\arctan\left(\frac{2+(\upalpha_0-\mu)\sigma}{\sigma\sqrt{\mu\upxi}}\right)-\pi\Bigg), \\
\mathfrak{B}(\sigma)	    &=& B(r(\sigma)) \nn \\
				&=& 1 - \dfrac{4\sigma(1-\sigma)\mu}{(2+\upalpha_0 \sigma)^2}.	
\eea
With these expressions, one derives from eqs.~\eqref{eq:F_of_sigma} and \eqref{eq:zeta}
\beq
{\cal K}(\sigma) = \sqrt{e^{\upupsilon(\sigma)} \mathfrak{B}(\sigma) }, \quad \zeta(\sigma) = \sqrt{\dfrac{e^{\upupsilon(\sigma)}}{\mathfrak{B}(\sigma) }}.
\eeq
At future null infinity $\sigma=0$ and the horizon $\sigma=1$, the function ${\cal K}(\sigma)$ assumes the values
\beq
{\cal K}(0)=1, \quad {\cal K}(1)={\cal K}_{\rm h} =  e^{ \sqrt{\frac{\mu}{\upxi}}\Bigg(\arctan\left(\frac{2+(\upalpha_0-\mu)}{\sqrt{\mu\upxi}}\right)-\dfrac{\pi}{2}\Bigg)}.
\eeq
Further metric components in the hyperboloidal line element \eqref{eq:conf_metric} requires the calculation of the dimensionless tortoise coordinate $x(\sigma)$. The singular terms follows from eqs.~\eqref{eq:x0} and \eqref{eq:x_horizons} and they read
\beq
x_0(\sigma) = \dfrac{\rh}{\lambda}\left( \dfrac{1}{\sigma} - (1+\mu) \ln \sigma \right), \quad x_{\rm h}(\sigma) = \dfrac{\rh}{\lambda} \dfrac{\ln(1-\sigma)}{{\cal K}_{\rm h}}.
\eeq

Similar to the previous section, the regular contribution to $x(\sigma)$ is non-trivial, and it is defined by the differential equation
\beq
\label{eq:xreg_BHHalo}
\dfrac{dx_{\rm reg}}{d\sigma} = \dfrac{r_{\rm h}}{\lambda}\dfrac{1}{\sigma^2 (1-\sigma)} \Bigg( -\dfrac{1}{{\cal K}(\sigma)} + \dfrac{\sigma^2}{{\cal K}_{\rm h}} + (1-\sigma) \bigg[1+(1+\mu)\sigma \bigg]\Bigg). 
\eeq
The limits $\sigma\rightarrow 0$ and $\sigma \rightarrow 1$ are regular in the above expression. The tortoise coordinate defines the metric function $p(\sigma)$, c.f.~eqs.~\eqref{eq:p_from_x} and \eqref{eq:def_p}
\beq
p(\sigma) = \dfrac{\rh}{\lambda} \sigma^2 (1-\sigma){\cal K}(\sigma).
\eeq

\begin{figure}[t!]
	\centering
	\includegraphics[scale=0.57]{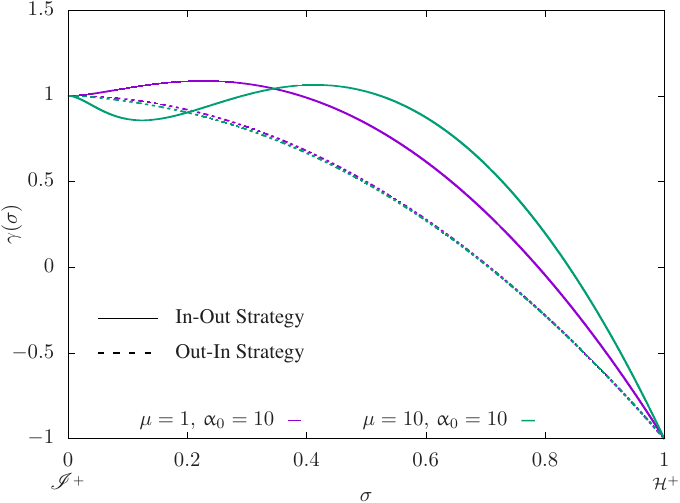}
	\includegraphics[scale=0.57]{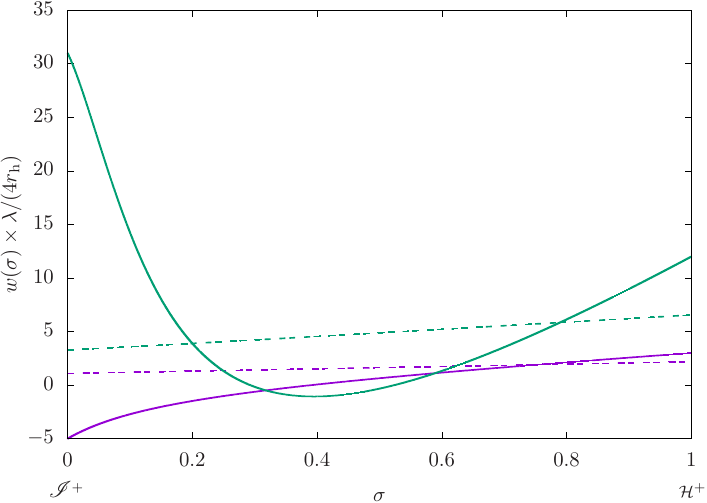}
	\caption{Metric functions $\gamma(\sigma)$ and $w(\gamma)$ for the black hole + halo spacetime in hyperboloidal slices. As in fig.~\ref{fig:gamma_w_highdim}, the in-out (solid lines) and out-int (dashed lines) strategies to construct the minimal gauge provide different results. Here, however conditions \eqref{eq:spacelike}, \eqref{eq:align_nullvectors_hrz} and \eqref{eq:align_nullvectors_scri} are not satisfied when using the in-out strategy, with $|\gamma|>1$ and $w<0$ within the domain $\sigma\in[0,1]$. The conditions hold when constructing the hyperboloidal slices with the out-in strategy.}
	\label{fig:gamma_w_BHHalo}
\end{figure}

With the expression for $x_0(\sigma)$, $x_{\rm h}(\sigma)$ and $x_{\rm reg}(\sigma)$, the height function in the minimal gauge assumes the structure given by eq.~\eqref{eq:H_MinGague_IO_OI}. Here, we also observe that the in-out and out-in strategies provide different solutions, with the remaining metric functions $\gamma(\sigma)$ and $w(\sigma)$ given by
\bea
\gamma^{\io}(\sigma) &=& -1 + 2(1-\sigma)\bigg(1 + \sigma(1+\mu) \bigg) {\cal K}(\sigma), \\
w^{\io}(\sigma) &=& \dfrac{4\rh}{\lambda \sigma^2}\bigg( 1+ \sigma (1+\mu) \bigg) \Bigg( 1 - (1-\sigma)\bigg[1+\sigma(1+\mu) \bigg] {\cal K}(\sigma)\Bigg),
\eea
or
\beq
\gamma^{\oi}(\sigma) = 1- \dfrac{2\sigma^2 {\cal K}(\sigma)}{{\cal K}_{\rm h}}, \quad w^{\oi}(\sigma) = \dfrac{4\rh}{\lambda (1-\sigma){\cal K}_{\rm h}} \bigg( 1 - \sigma^2 \dfrac{{\cal K}(\sigma)}{{\cal K}_{\rm h}}\bigg).
\eeq
Contrary to the results in sec.~\ref{sec:HigDimSch}, not only are the expressions for the in-out strategy lengthier than then ones derived with the out-in strategy, but they may also violate conditions \eqref{eq:spacelike}, \eqref{eq:align_nullvectors_hrz} and \eqref{eq:align_nullvectors_scri}. The violation is observed in fig.~\ref{fig:gamma_w_BHHalo}, where the functions assumes values $|\gamma(\sigma)|>1$ and $w(\sigma)<1$ within the exterior black-hole region $\sigma\in[0,1]$ (in-out strategy --- solid lines). On the other hand, the conditions \eqref{eq:spacelike}, \eqref{eq:align_nullvectors_hrz} and \eqref{eq:align_nullvectors_scri} are always satisfied when using the out-in strategy (dashed lines). Fig.~\ref{fig:gamma_w_BHHalo} shows results for the parameters $(\mu, \upalpha_0)=(1,10)$ (purple) and $(\mu, \upalpha_0)=(10,10)$ (green). We have empirically observed that conditions \eqref{eq:spacelike}, \eqref{eq:align_nullvectors_hrz} and \eqref{eq:align_nullvectors_scri} holds in the out-in strategy for all studies and experiments developed in refs.~\cite{Cardoso:2021wlq,Cardoso:2022whc}.

\subsection{Reisnner-Nordstr\"om-de Sitter spacetimes}
With finish the result section with the Reisnner-Nordstr\"om-de Sitter spacetime. Contrary to the previous examples, this solution has a richer structure of horizons. Here, future infinity $\scri^+$ is still located at $\sigma=0$, but this surface is space-like. The causal domain for the exterior black hole region is restricted between the event horizon ${\cal H}^+$ and the cosmological horizon ${\cal H}^+_{\Lambda}$. Besides, the interior black hole region also display a Cauchy horizon ${\cal C}^+$ whenever the black hole is charged.

The Reisnner-Nordstr\"om-de Sitter spacetime is given by the line element \eqref{eq:metric_tr} with metric functions
\beq
a(r) = b(r) = 1 - \dfrac{2M}{r} + \dfrac{Q^2}{r^2} - \dfrac{\Lambda}{3}r^2.
\eeq
Instead of parametrising the spacetime by its mass $M$, charge $Q$ and cosmological constant $\Lambda$, we opt to express the metric functions in terms of the radial coordinate values for the event horizon $\rh$, Cauchy horizon $r_{\rm C}$ and cosmological horizon $r_\Lambda$ via
\beq
\label{eq:RNdSHorizons}
a(r) = b(r) =- r^2\dfrac{\Lambda}{3}\left( 1- \dfrac{\rh}{r}\right)\left( 1- \dfrac{r_{\rm C}}{r}\right)\left( 1-\dfrac{r_\Lambda}{r} \right)\left( 1  - \dfrac{r_o}{r} \right).
\eeq
Eq.~\eqref{eq:RNdSHorizons} has precisely the form presented in eq.~\eqref{eq:func_f_horizons}, but the root $r_o = - \left( \rh + r_{\rm C} + r_\Lambda\right)$ is non-physical as it assumes negative values. With the radial compactification \eqref{eq:r_of_sigma}, the horizons and the negative root are mapped into
\beq
\sigmah = 1, \quad \sigma_{\rm C}= \rh/r_{\rm C}, \quad \sigma_{\Lambda}= \rh/r_{\Lambda}, \quad \sigma_{o}= \rh/r_{o}.
\eeq
Eq.~\eqref{eq:x_horizons} applied to each one of the $n_{\rm h}= 4$ roots of \eqref{eq:RNdSHorizons} yields the dimensionless tortoise coordinate
\beq
\label{eq:x_RNdS}
x(\sigma) = x_{\rm h}(\sigma) + x_{\rm C}(\sigma) + x_{\Lambda}(\sigma) + x_{o}(\sigma),
\eeq
with
\bea
x_{\rm h}(\sigma) &=&- \dfrac{3\, \sigmah}{\lambda r_h \Lambda}\dfrac{\ln(\sigmah-\sigma)}{(1 - \sigmah/\sigma_{\rm C})(1 - \sigmah/\sigma_\Lambda)(1 - \sigmah/\sigma_o)}, \\
x_{\rm C}(\sigma) &=&- \dfrac{3 \, \sigma_{\rm C}}{\lambda r_h \Lambda}\dfrac{\ln(\sigma_{\rm C}-\sigma)}{(1 - \sigma_{\rm C}/\sigmah)(1 - \sigma_{\rm C}/\sigma_\Lambda)(1 - \sigma_{\rm C}/\sigma_o)}, \\
x_{\Lambda}(\sigma) &=& - \dfrac{3 \, \sigma_\Lambda}{\lambda r_h \Lambda}\dfrac{\ln(\sigma-\sigma_{\Lambda})}{(1 - \sigma_{\Lambda}/\sigmah)(1 -  \sigma_{\Lambda}/\sigma_{\rm C})(1 -  \sigma_{\Lambda}/\sigma_o)}, \\
x_{o}(\sigma)&=& - \dfrac{3 \sigma_o}{\lambda r_h \Lambda}\dfrac{\ln(\sigma-\sigma_{o})}{(1 - \sigma_{o}/\sigmah)(1 -  \sigma_{o}/\sigma_{\rm C})(1 -  \sigma_o/\sigma_{\Lambda})}.
\eea
One can verify that eq.~\eqref{eq:x_RNdS} reproduces the behaviour \eqref{eq:x0_dS} as $\sigma\rightarrow 0$. Besides, if we apply the complete expression ~\eqref{eq:x_RNdS} into eq.~\eqref{eq:dxreg_dsigma}, one observes that $x'_{\rm reg} =0$. However, it is crucial to keep in mind that the contribution from $x_o(\sigma)$ comes from the non-physical root $r_o<0$. Thus, we must identify the tortoise coordinate's regular part direct as $x_{\rm reg}(\sigma):=x_o(\sigma)$.

The results from sec.~\ref{sec:MinGauge_Height} shows that the minimal gauge height function in the in-out strategy follows from just changing the sign of the cosmological term $x_{\Lambda}(\sigma)$ within the expression for the dimensionless tortoise coordinate, i.e.
\beq
\label{eq:H_RNdS_inout}
H^{\rm in-out}(\sigma) = x_{\rm h}(\sigma) + x_{\rm C}(\sigma) - x_{\Lambda}(\sigma) + x_{o}(\sigma).
\eeq
Having identified $x_o$ with the regular piece in $x(\sigma)$, the minimal gauge height function in the out-in strategy follows from also changing the sign of $x_{\rm reg}$, thus 
\beq
\label{eq:H_RNdS_outin}
H^{\rm out-in}(\sigma) = x_{\rm h}(\sigma) + x_{\rm C}(\sigma) - x_{\Lambda}(\sigma) - x_{o}(\sigma).
\eeq
The metric functions $p(\sigma), \gamma(\sigma)$ and $w(\sigma)$ are directly derived from eqs.~\eqref{eq:x_RNdS}-\eqref{eq:H_RNdS_outin} with the help of eqs.~\eqref{eq:p_from_x}, \eqref{eq:def_gamma} and \eqref{eq:def_w}, respectively. The result is rather lengthy and we omit their explicit expressions here, noting that any mathematical computation program calculates the results effortlessly. As in the previous sections, we study the behaviour of the function $w(\sigma)$ for several parameters of the solution to ensure the hyperboloidal surfaces are space-like in the domains of interest.

\begin{figure}[t!]
	\centering
	\includegraphics[scale=0.55]{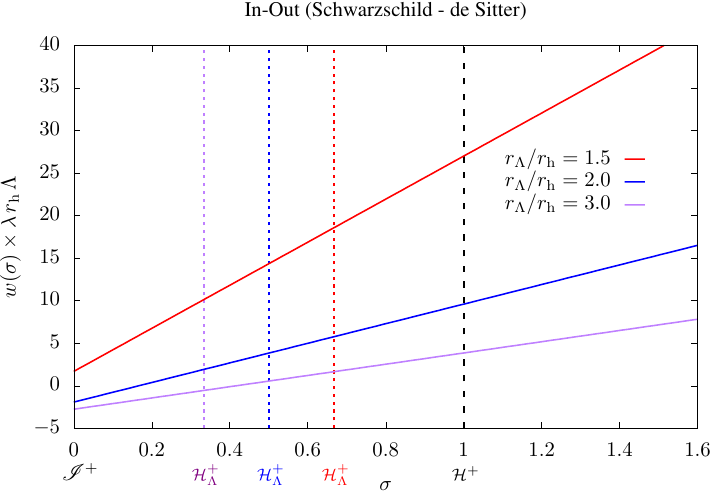}
	\includegraphics[scale=0.55]{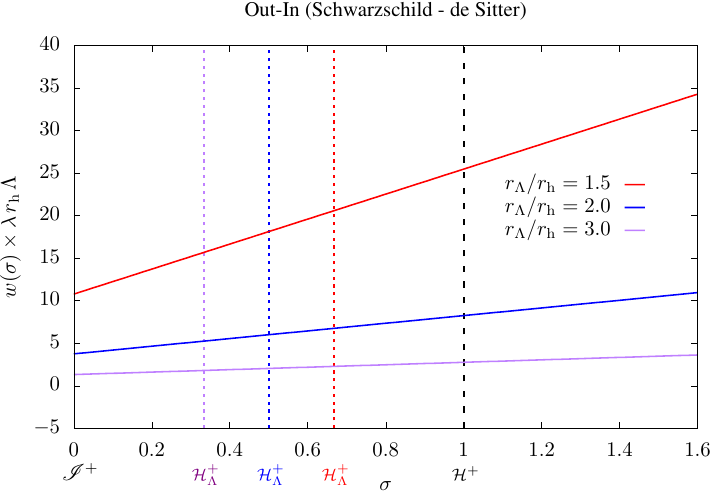}
	\caption{Metric function  $w(\gamma)$ for the Schwarzschild-de Sitter spacetime in the hyperboloidal minimal gauge. {\em Left Panel:} With the in-out strategy, the function $w(\gamma)$ assumes negative values as the cosmological horizon increases. For $r_\Lambda = 3\rh/2$ (red), $w(\gamma)$ is still positive in the entire domain $\sigma[0,\infty)$. For $r_\Lambda = 2\rh$ (blue), the hyperboloidal time surfaces are spacelike in the causal domain of the exterior black hole region $\sigma\in[\sigma_\Lambda, \sigmah]$, but $w(\sigma)$ becomes negative as one extends the surface from $\sigma_\Lambda$ towards $\scri^+$. For $r_{\Lambda} = 3 \rh$ (purple), $w(\sigma)<0$ already in the region $\sigma\in[\sigma_\Lambda, \sigmah]$. {\em Right Panel:} With the out-in strategy hyperboloidal time surfaces are always spacelike: $w(\gamma)>0$ from the (spacelike) future infinity $\scri^+$ ($\sigma=0$) down to the physical singularity $(\sigma\rightarrow \infty$), regardless of the values for the cosmological horizon.}
	\label{fig:SdS}
\end{figure}

We begin by considering the Schwarzschild-de Sitter solution, by setting the Cauchy horizon parameter $r_{\rm C} = 0 \Longrightarrow \sigma_{\rm C}\rightarrow \infty$. The left panel of fig. \ref{fig:SdS} brings the function $w(\sigma)$ for the in-out strategy. The important behaviours are captured by three representative parameter. For relative small values of $r_{\Lambda}/r_h$, e.g. $r_{\Lambda} = 3 \rh/2$ (red), $w(\sigma)>0$ in the domain $\sigma\in[0,\infty)$. As one increases the cosmological horizon, e.g. $r_{\Lambda} = 2 \rh$ (blue), the function $w(\sigma)$ remains positive in the causal domain of the exterior black hole region $\sigma\in[\sigma_\Lambda, \sigmah]$. However, the hyperboloidal time surface changes property as one extends the solution from $\sigma_\Lambda$ inside, towards $\scri^+$. By pushing the cosmological horizon to higher values, e.g. $r_{\Lambda} = 3 \rh$ (purple), $w(\sigma)$ changes sign already in the domain $\sigma\in[\sigma_\Lambda, \sigmah]$. We do not encounter these properties with the out-in strategy, and $w(\sigma)>0$ in the entire domain $\sigma[0,\infty)$, regardless of the value for the cosmological horizon (see right panel of fig.~\ref{fig:SdS}). As in the asymptotic flat case from sec.~\ref{sec:trumpet}, the hyperboloidal foliation is valid from the (spacelike) future infinity $\sigma=0$ up to the (spacelike) singularity $\sigma\rightarrow \infty$.

\begin{figure}[b!]
	\centering
	\includegraphics[scale=0.55]{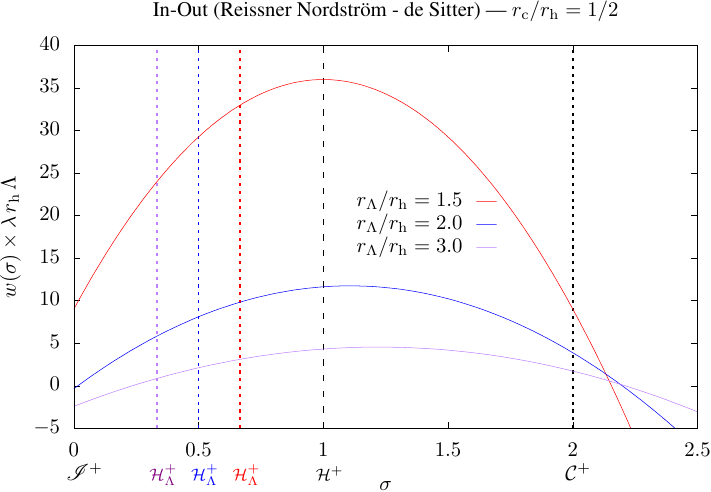}
	\includegraphics[scale=0.55]{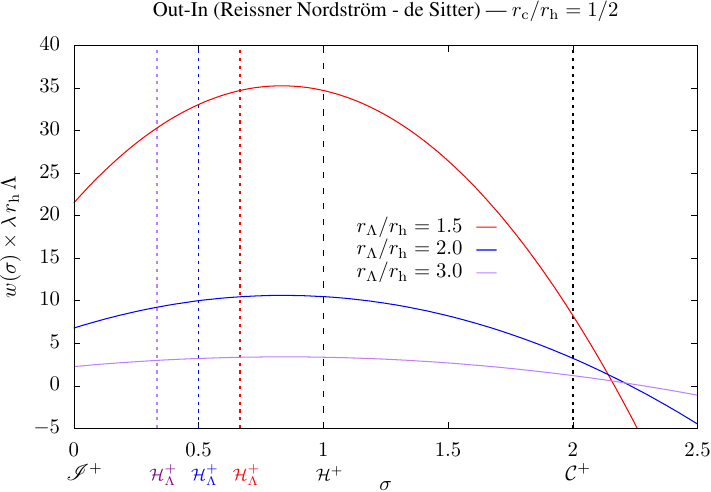}
	\caption{
Metric function  $w(\gamma)$ for the Reissner Nordstr\"om-de Sitter spacetime in the hyperboloidal minimal gauge. Example for Cauchy horizon value $r_{\rm C} = \rh/2$, but results are qualitatively the same up to the extreme configuration $r_{\rm C} = \rh$. In the domain $\sigma\in[0,\sigma_{\rm C}]$, the property of the hyperboloidal time surfaces are the same as in the uncharged case (fig.~\ref{fig:SdS}):
 in the in-out strategy (left panel), $w(\sigma)$ assumes negative values as one increases the cosmological horizon $r_{\Lambda}$, whereas the hyperboloidal surfaces remain spacelike in the entire domain within the out-in strategy (right panel). In the domain $\sigma\in[\sigma_{\rm C},\infty)$, as one crosses the Cauchy horizon towards the (timelike) singularity, the hyperboloidal surfaces in the minimal gauge become unavoidably timelike.
}
	\label{fig:RNdS}
\end{figure}

Fig.~\ref{fig:RNdS} shows the results for the Reisnner N\"ordstron-de Sitter solution. In the example, we fix the Cauchy horizon to $r_{\rm C} = \rh/2$, but the results are qualitatively the same even in the extremal case $r_{\rm C} = \rh$. In the domain $\sigma\in [0,\sigma_{\rm C}]$, we observe exactly the same properties as in the Schwarzschild-de Sitter case: The function $w(\sigma)$ develop regions with negative values for in-out strategy as one increases the cosmological horizon $r_{\Lambda}$, whereas the hyperboloidal surfaces remain spacelike in the entire domain within the out-in strategy. There is, however, one significant different behaviour when black holes are charged. The hyperboloidal surfaces in the minimal gauge become unavoidably timelike, once one crosses the Cauchy horizon towards the singularity. This is a direct consequence of the physical singularity being itself timelike in Reisnner N\"ordstron spacetimes. 

\section{Conclusion}
This work presents a comprehensive discussion on the conformal compactification of spherically symmetric spacetimes via the hyperboloidal approach. The hyperboloidal foliations consists of spacelike hypersurfaces with asymptotically hyperbolic geometry. On a fixed background,  these surfaces emerges as the level set of a time coordinate $\tau=$ constant, which provides a notion for accessing spacetime null surfaces (such as the black-hole horizon and future null infinity) "at the same time".  

The choice for a particular hyperboloidal foliation is not unique, and we present the formalism for a generic spherically symmetric spacetime without fixing ourselves to a particular gauge, at first. In particular, starting from the most generic line element in the usual Scharzschild-like coordinates $(t,r,\theta,\varphi)$, we introduce the transformation to hyperboloidal coordinates $(\tau, \sigma,\theta,\varphi)$ and discuss the geometrical and physical interpretation to all functions in the hyperboloidal line element \eqref{eq:conf_metric}. For instance, eq.~\eqref{eq:BondiMass} allows us to calculate the Trautman-Bondi mass directly from the hyperboloidal quantities.

The hyperboloidal approach is a powerful tool for studying wave propagation problems and it has proved its value in the field of black-hole perturbation theory. Thus, this work reviews and applies the generic hyperboloidal framework to a wide class of wave equations relevant to the linear regime of general relativity. The framework provides the starting point for studies of binary black holes dynamics via the the hyperboloidal approach where the linear regime is applicable: the inspiral of objects with extreme mass-ratios and the ring-down. 

Even though the discussion presented is generic enough to comprise a large class of scenarios relevant to astrophysical studies, it is restricted to problems in which the dynamics are reducible to one wave equation for a single master function.  For instance, a generalisation of the formalism is required when using hyperboloidal tools to studies of EMRI's via the self-force programme in the Lorenz gauge\cite{Akcay:2010dx,Akcay:2013wfa,Osburn:2022bby}. Further development is also needed for scenarios with matter distributions, where the wave equations for density fluctuations couple with the gravitational waves degrees of freedom\cite{Allen:1997xj,Cardoso:2022whc}. Overall, we do not anticipate any conceptual difficulty in extending the framework to a system of coupled wave equations when they all share the same characteristic speed of propagation, such as the formulations common to the gravitational self-force programme \cite{Akcay:2010dx,Akcay:2013wfa,Osburn:2022bby,Durkan:2022fvm,Dolan:2023enf}. The framework is also applicable to a system of coupled wave equations with varying propagation speeds due to matter distributions, but a more careful analysis of the solutions' regularity properties might be needed.

Among the possible choices for hyperboloidal foliations, this work re-visits the so-called minimal gauge. This class of gauges provides the minimal structure for the coordinate degrees of freedom, so that the surfaces of $\tau=$ constant are hyperboloidal. The procedure for its construction was initially identified by Marcus Ansorg in refs.~\cite{Schinkel:2013tka,Schinkel:2013zm,PanossoMacedo:2014dnr} and formalised in refs.~\cite{Ansorg:2016ztf,PanossoMacedo:2018hab,PanossoMacedo:2019npm}. This work develops further this technique to make it applicable to any spherically symmetric spacetime. 

The most important result is that the height function in the minimal gauge follows directly from the tortoise coordinate by a simple change of sign in the terms singular at future null infinity (or cosmological horizon for asymptotically de-Sitter spacetimes). Even though this results provides a simple algorithm for constructing the height function in the minimal gauge, one must ensure that the resulting $\tau=$ constant surfaces are indeed hyperboloidal and satisfy conditions \eqref{eq:spacelike}-\eqref{eq:align_nullvectors_scri}. 

More specifically, this works shows that the conceptual strategy for constructing the minimal gauge height function can be cast into two alternative schemes: (i) the in-out strategy and (ii) the out-in strategy. The former considers first ingoing null geodesics, to ensure that black-hole horizons are intersected by the time surfaces. Then, it integrates outgoing null geodesics asymptotically around future null infinity (or cosmological horizons). The latter considers first outgoing null geodesics, to ensure that future null infinity (or cosmological horizons) are intersected. Then, it integrates ingoing null geodesics around the horizons. 

These two strategies are completely equivalent for a large class of spacetimes, such as Schwarzschild, Reisnner-N\"ordstrom, extending up to Kerr. However, they may differ for more generic solutions. In this work, we identify the reason why the in-out and out-in strategies may not lead to the same results. The answer lies in the structure of the tortoise coordinate. We express the tortoise coordinates as the sum of terms singular at future null infinity and horizons, plus a regular term. For solutions in which the regular term is a constant (usually zero), the in-out and out-in strategies provide the same hyperboloidal foliations in the minimal gauge. For solutions with a non-trivial regular piece, these strategies yield different results, as expressed in eq.~\eqref{eq:Hinout-Houtin}. The preference for one strategy over the other is based first on mathematical grounds, i.e., if conditions \eqref{eq:spacelike}-\eqref{eq:align_nullvectors_scri} are violated and the resulting $\tau=$ constant surfaces are not spacelike, and second on practical terms, i.e., if one approach gives simpler expressions than the other.   

Finally, we apply the formalism to the following spacetimes: Schwarzschild, Schwarschild-Tangherlini, black-hole + halo and Reisnner-N\"ostroim-de Sitter. Each one of these examples discusses one particular aspect of the hyperboloidal minimal gauge, such as the choice for compactification function \eqref{eq:rho_mingauge} and the outcomes for the in-out and out-in strategy. Most importantly, the example in sec.~\ref{sec:BHHalo} shows that the out-in strategy is preferable. These heuristic results motivate us to conjecture that hyperboloidal slices in the minimal gauge are always available via the out-in strategy, but a formal proof goes beyond the scope of this work.

\section*{Acknowledgments}
The author would like to thank Anil Zenginoglu, David Hilditch and Alex Va\~no-Vi\~nuales for valuable discussions. Also, the author thanks Juan Valiente Kroon, Grigalus Tayjanskas and the Royal Society for the invitation and financial support to join the scientific meeting "At the interface of asymptotic, conformal methods and analysis in general relativity". This work was financed by the VILLUM Foundation (grant no. VIL37766), the DNRF Chair program (grant no. DNRF162) by the Danish National Research Foundation, and the European Union’s H2020 ERC Advanced Grant "Black holes: gravitational engines of discovery" grant agreement no. Gravitas–101052587. Views and opinions expressed are however those of the author only and do not necessarily reflect those of the European Union or the European Research Council. Neither the European Union nor the granting authority can be held responsible for them.

\section*{Bibliography}
\bibliographystyle{unsrt.bst}
\bibliography{bibitems}

\end{document}